\documentclass[12pt,preprint]{aastex}
\usepackage{epsf}

\newenvironment{inlinefigure}{
\def\@captype{figure}
\noindent\begin{minipage}{0.999\linewidth}\begin{center}}
{\end{center}\end{minipage}\smallskip}

\slugcomment{ApJ, in press}
\shorttitle{Arcs in the HST/WFPC2 Archive}
\shortauthors{Sand et al.}

\begin{document}
\title{A Systematic Search for Gravitationally-Lensed Arcs
in the Hubble Space Telescope WFPC2 Archive}

\author{David J. Sand, Tommaso Treu\altaffilmark{1,}\altaffilmark{2}, Richard S. Ellis and Graham P. Smith}
\affil{California Institute of Technology,
Astronomy, mailcode 105--24, Pasadena, CA 91125}
\email{djs@astro.caltech.edu}
\altaffiltext{1}{Hubble Fellow}
\altaffiltext{2}{Current Address: Department of Physics \& Astronomy, UCLA, Box 951547, Los Angeles, CA 90095-1547}

\begin{abstract}

We present the results of a systematic search for
gravitationally-lensed arcs in clusters of galaxies located in the
Hubble Space Telescope Wide Field and Planetary Camera 2 data
archive. By carefully examining the images of 128 clusters we have
located 12 candidate radial arcs and 104 tangential arcs, each of
whose length to width ratio exceeds 7. In addition, 24 other radial
arc candidates were identified with a length to width ratio less than
7.  Keck spectroscopy of 17 candidate radial arcs suggests that
contamination of the radial arc sample from non-lensed objects is
$\sim$30-50\%. With our catalog, we explore the practicality of using
the number ratio of radial to tangential arcs as a statistical measure
of the slope $\beta$ of the dark matter distribution in cluster cores
(where $\rho_{DM}\propto r^{-\beta}$ at small radii). Despite the
heterogeneous nature of the cluster sample, we demonstrate that this
abundance ratio is fairly constant across various cluster subsamples
partitioned according to X-ray luminosity and optical survey depth. We
develop the necessary formalism to interpret this ratio in the context
of two-component mass models for cluster cores.  Although the arc
statistics in our survey are consistent with a range of density
profiles -- $\beta \lesssim $1.6 depending on various assumptions, we
show that one of the prime limiting factors is the distribution of
stellar masses for the brightest cluster galaxies. We discuss the
prospects for improving the observational constraints and thereby
providing a reliable statistical constraint on cluster dark matter
profiles on $\lesssim$100 kpc scales.

\end{abstract}
\keywords{gravitational lensing: radial arcs ---  cD }

\section{Introduction}

The imaging cameras on Hubble Space Telescope ($HST$) provide a
valuable resource for studies of gravitational lensing. For example,
the improved image quality compared to ground-based telescopes has
enabled the morphological recognition of tangential arcs (e.g.~Smail
et al 1996; Kneib et al 1996; Gioia et al. 1998). The analysis of such
arcs has led to detailed mass models of great utility both in
determining dark and baryonic mass distributions (e.g.~Kneib et al
2003; Gavazzi et al. 2003; Smith et al. 2005) and in the study of
highly magnified distant galaxies (e.g.~Franx et al. 1997; Seitz et
al. 1998; Pettini et al. 2000; Smith et al. 2002; Swinbank et
al. 2003; Ellis et al 2001; Santos et al 2004; Kneib et al 2004).

HST images have also been invaluable in studying {\it radial}
gravitationally-lensed arcs (e.g.~Gioia et al. 1998; Smith et
al. 2001; Sand, Treu \& Ellis 2002; Sand et al. 2004). These arcs are
often embedded in the envelope of the central luminous cluster galaxy
and thus a high angular resolution is essential to uncover their
presence. Radial arcs straddle the inner critical line whose location
has long been known to provide a valuable constraint on the form of
the mass profile on $\lesssim$100 kpc scales (e.g.~Fort et al 1992;
Miralda-Escud\'e 1993; Miralda-Escud\'e 1995; Bartelmann 1996;
Williams et al. 1999; Meneghetti et al. 2001).

A long-standing field of inquiry has been the comparison of
theoretical predictions and ground-based observations of the {\it
abundance of arcs} for example as a constraint on
cosmology. Bartelmann et al. (1998) originally found that the number
of strongly-lensed arcs greatly exceeds that expected from
$\Lambda$CDM simulations, preferring instead an open CDM
cosmology. Various systematic effects have been proposed to explain
the apparent excess including cluster substructure (e.g. Flores,
Maller, \& Primack 2000; Torri et al. 2004), the influence of the
brightest cluster galaxy (BCG; Meneghetti, Bartelmann, \& Moscardini
2003), and uncertainties in the background redshift distribution of
lensed sources (Dalal, Holder, \& Hennawi 2004; Wambsganss, Bode, \&
Ostriker 2004).  Many of these effects can be calibrated through high
resolution simulations of galaxy clusters and accurate background
redshift distributions based on photometric redshift surveys. Indeed,
several recent articles reconcile the expected number of gravitational
arcs in a $\Lambda$CDM universe with observations (Dalal et al. 2004;
Wambsganass et al. 2003; Oguri et al. 2003) and attention is now
focusing on how to use such observations to constrain dark energy
models (e.g.~Meneghetti et al. 2004; Dalal, Hennawi \& Bode 2004).

It is also possible to use arc statistics to constrain the {\it
central density profiles} of clusters (e.g. Wyithe, Turner, \& Spergel
2001), thereby testing the prediction that CDM halos have profiles
steeper than $\rho\propto r^{-1.0}$ (e.g. Navarro, Frenk, \& White
1997; Moore et al. 1998; Power et al. 2003; Fukushige, Kawaii, \&
Makino 2004; Tasitsiomi et al. 2004; Diemand et al. 2004). These
analyses are subject to uncertainties and systematics similar to those
discussed above. 

To date, several ground-based optical surveys have been used for
statistical studies of gravitational arcs (e.g. Le Fevre et al.  1994;
Luppino et al.  1999; Zaritsky \& Gonzalez 2003; Gladders et
al. 2003).  Despite different cluster selection and redshift criteria,
these surveys have measured roughly comparable giant tangential
gravitational arc incidences which have guided theoretical
understanding of the processes responsible for strong lensing on the
galaxy cluster scale.  Given typical ground based seeing, however,
these are of marginal utility in searches for radially-elongated thin,
faint arcs buried in the halos of bright cluster galaxies.

The primary goal of this paper is to compile a list of the
gravitationally lensed arcs found in the HST/WFPC2 archive and to
explore the feasibility of using the number ratio of radial to
tangential arcs as a means of constraining the inner density profiles
in cluster cores.  Molikawa \& Hattori (2001) have shown that the
abundance ratio of radial to tangential arcs is sensitive to the mean
density profile of the cluster sample. Oguri, Taruya, \& Suto (2001)
studied the predicted ratio of radial to tangential arcs as a function
of not only the inner dark matter density slope, but the concentration
parameter, $c$, of the halos as well.  Oguri (2002; hereafter O02; see
also Keeton 2001) has suggested that the various systematics which
effect the cross section for lensing are significantly reduced when
considering the number ratio of radial to tangential arcs rather than
their absolute number.  To constrain the dark matter density profile,
we adopt the methodology presented by O02, extending their technique
to include a second mass component arising from the central cluster
galaxy.  Our analysis is intended to complement studies of the DM
density profile in clusters performed on individual systems
(e.g. Kneib et al. 2003; Smith et al. 2001; Gavazzi et al. 2003; Sand
et al. 2002; Sand et al. 2004; Buote et al. 2004; Lokas \& Mamon 2003;
Kelson et al. 2002) and through other statistical techniques (e.g.~van
der Marel 2000; Dahle et al. 2003; Mahdavi \& Geller 2004).

A plan of the paper follows.  In \S 2 we discuss the archival sample
of clusters, how representative sub-samples can be defined for later
analyses and describe our reduction procedure. In \S 3 we describe the
procedures we adopted for identifying lensed arcs and how they are
characterized by their length-to-width ratio. We also present new
follow-up spectroscopy for several candidate radial arcs as a means of
estimating the likely contamination by other sources (e.g. foreground
galaxies).  In \S 4 we present our methodology for calculating the
expected radial to tangential arc number ratio and discuss the various
assumptions and their limitations. In \S 5 we derive constraints on
the inner DM density slope and discuss our results.  In \S 6 we
summarize and discuss future prospects for improving the
constraints. An Appendix describes and presents the cluster catalog,
arc catalog and finding charts for the newly-located radial arcs.

Throughout this paper, we adopt $r$ as the radial coordinate in
3-D space and $R$ as the radial coordinate in 2-D projected space.
We assume H$_0$=65~km s$^{-1}$Mpc$^{-1}$, $\Omega_m=0.3$ and
$\Omega_{\Lambda}$=0.7.

\section{Cluster Selection \label{sec:clussec}}

The Hubble Space Telescope data archive is now sufficiently extensive
to provide the basis for a search for gravitational arcs in galaxy
clusters. In this work, we restricted our search to images taken with
the Wide Field Planetary Camera 2 (WFPC2). Exploitation of the
archival set of images taken with the more recently-installed and
superior Advanced Camera for Surveys (ACS) is left for a future study.

As the overarching goal is to identify all tangentially and
radially-elongated gravitational arcs, regardless of any preordained
intrinsic property of the galaxy cluster, fairly liberal criteria were
used for selecting observations from the archive. Only clusters of
known redshift with $0.1<z<0.8$ were considered. We stipulated that
images of the cluster had to be available in one or more of the
following broad band filters: F450W, F555W, F606W, F675W, F702W and
F814W.  Procedurally, an abstract search was done on the HST archive
and proposals containing the words ``galaxy'' and ``cluster'' or
``group'' were flagged.  This initial list of abstracts was pared by
inspection, so that only data for those proposals directed at galaxy
clusters or groups were requested.  All of the data from this edited
abstract list was requested if they satisfied the camera, filter and
redshift requirements.  This search technique ensures only programs
deliberately targetting clusters are examined.

The resulting cluster catalog is listed in Table~\ref{fig:clustable},
in the Appendix.  The total sample includes 128 galaxy clusters and a
histogram of the redshift distribution is shown in
Figure~\ref{fig:cluszfig}.

\subsection{Uniform Cluster Subsamples \label{sec:subsample}}

The resulting cluster sample is heterogenous with factors such as
exposure time, redshift, richness/mass/X-ray luminosity and filter
choice all affecting the sensitivity to gravitational arcs. Although
this may not seriously affect our goal of measuring the abundance
ratio of radial to tangential arcs and constraining the inner slope of
the DM density profile (see O02; the arc ratio is relatively robust
with respect to cluster mass and observational selection effects), it
is helpful to consider the possibility of partitioning the large
sample into more complete subsets for later use. Membership of each
cluster in the various sub-samples introduced below is indicated in
Table~\ref{fig:clustable}.

X-ray selected sub-samples will be beneficial given the
correlation between cluster mass and X-ray luminosity. We define
two in particular that have been discussed in the literature and
which link directly to specific HST programs.

1.{\it Smith sample:} This sample follows the work of Smith et
al. (2001,2002a,2002b,2003,2005) and includes 10 clusters.  Clusters
in this sample are X-ray luminous ($L_{\rm X}>8\times10^{44}$ ergs
$s^{-1}$; 0.1-2.4 keV; Ebeling et al. 1996) and lie in the redshift
range $0.17<z<0.25$.

2.{\it EMSS sample:} Another X-ray sub-sample can be drawn from the
EMSS cluster survey (0.3-3.5 keV; limiting sensitivity of
$5\times10^{-14}$ ergs $cm^{2}$ $s^{-1}$; Henry et al. 1992). Of the
93 clusters identified by Henry et al. twelve have been imaged with
HST/WFPC2. Two previous gravitational arc searches were conducted with
a subsample of this kind using ground-based images (Le Fevre et al
1994; Luppino et al. 1999).

3. {\it Edge sample:} This refers to a sample of clusters whose
imaging was conducted to a uniform depth (although the clusters are
not all the same redshift). Such a sample should, broadly speaking,
pick out all lensed features to a certain surface brightness
threshold. Two large SNAPSHOT programs are prominent in this respect:
PID 8301 and 8719 (PI: Edge) which image together 44 $z>$0.1 clusters
in the F606W filter with exposure times between 0.6 and 1.0
ks. According to the HST proposals, this program sought to understand
the morphological properties of central cluster galaxies.  Clusters
were selected from the Brightest Cluster Sample (BCS -- Ebeling et
al.\ 1998) for which optical spectra of the central cluster galaxies
are available (Crawford et al.\ 1999).  The primary sample was
selected from those BCS clusters hosting a BCG with optical line
emission.  A secondary control sample of BCS clusters that do not host
an optical line emitting BCG was also selected to span the same range
in redshift and X--ray luminosity as the primary sample (Edge, priv.\
comm.).  Optical line emission from BCGs is one of the least ambiguous
indicators of clusters for which the central cooling timescale is less
than the Hubble time (Crawford et al.\ 1999).  These "cooling flow"
clusters are also typically classified as relaxed clusters (e.g.\
Smith et al.\ 2005).  The clusters in the control sample were also
selected to appear relaxed at optical and X--ray wavelengths.  While
there are undoubtedly some exceptions (e.g.\ Edge et al.\ 2003), the
"Edge sample" is likely dominated by relaxed clusters.

\subsection{WFPC2 Data Reduction \label{sec:redux}}

\sloppypar Although our cluster sample is drawn from the HST/WFPC2
archive, the various goals of each original program means the
observing strategy varied from case to case. Fortunately, however,
there are only two basic approaches to taking the observations. The
first includes those CRSPLIT or SNAPSHOT observations in which two or
more non-dithered exposures were taken. The second refers to the case
where two or more dithered (either with integer or half-integer pixel
offsets) exposures were taken either to enhance the sampling of the
WFPC2 point spread function or for better cosmic ray removal (or
both). A standard data reduction script was written for each of these
cases and is described here briefly.

In the SNAPSHOT case, cosmic ray rejection was first performed on
each individual exposure using the {\sc iraf} task {\sc lacosmic}
(van Dokkum 2001). The cleaned images were then combined with the
task {\sc crrej}, which also served to remove residual cosmic ray
hits. Background counts subtracted from each of the WFPC2 chips
were noted and used in later photometric calculations. The WFPC2
chips were combined using the {\sc iraf} task {\sc wmosaic}.

For the multiple dithered exposures, the data were reduced using the
{\sc iraf} package DRIZZLE (Fruchter \& Hook 2002) with a fixed
parameter set. In particular, the final pixel size (represented by the
{\sc drizzle.scale} parameter) was set to 0.5 resulting in a pixel
size half that of the original image. The drizzled ``drop'' size
(represented by the {\sc drizzle.pixfrac} parameter) was set to 0.8
regardless of the observational program.  The final images of the
WFPC2 chips were combined using the {\sc iraf} task {\sc gprep}. The
sky background determined for each WFPC2 chip was again noted for
later photometric use.

To aid in locating radial arcs, which are often buried in the halo of
the BCG, we also examined images after subtracting the (assumed
symmetrical) light of the most luminous galaxies (usually but not
always just the central member). To do this we employed the {\sc iraf}
task {\sc ellipse} allowing both the position angle and ellipticity of
the fitted isophotes to vary as a function of semi-major axis. We
discuss in \S 3.1, \S 3.2, \& \S 3.3 how the galaxy subtraction and
residuals might affect arc identification, photometry and derived
length to width ratios, respectively.

\section{The Arc Sample}

In this section we discuss how the sample of tangential and radial
arcs were identified in a consistent manner from the reduced data. The
resulting catalog is presented in Table~\ref{fig:arctable} in the
Appendix.

\subsection{Arc Identification \label{sec:arcid}}

Each mosaiced image was visually examined for lensed features by one
of the authors (DJS), both in its original and bright
galaxy-subtracted incarnations.  A candidate gravitational arc was
designated according to one of two categories: tangential or radial
arc. The distinction between tangential and radial arcs is determined
by the arc orientation with respect to the cluster center (assumed to
be roughly coincident with the dominant BCG) and is rarely ambiguous,
even in bi- or multi-modal clusters.  Twenty-five of the HST clusters
were also examined by one other author (TT).  Within this subsample,
DJS found 2 radial and 37 tangential arcs with a $L/W>$7 (see
\S\ref{sec:method} for justification of this $L/W$ criterion), while
TT found 2 radial and 41 tangential arcs, resulting in arc number
ratios of $0.054^{+0.077}_{-0.036}$ and $0.049^{+0.069}_{-0.032}$,
respectively.  Thirty-six of the tangential arcs and both of the
radial arcs were in common between the two samples.  Given the
consistency between these measurements, we conclude that our results
are not sensitive to the person doing the identification.

A serious concern given our motivation to measure the {\it ratio} of
the radial to tangential arcs, is the likelihood that radial arcs are
harder to locate in the noisier region underneath the envelope of the
brightest cluster galaxy. Taking the ten cluster Smith sample (see
\S~\ref{sec:subsample}) as an example, we find the rms background
noise to be 1.5 to 3 times higher in the central regions after galaxy
subtraction than in the periphery of the WFPC2 fields appropriate for
the identification of tangential arcs. To investigate the bias this
might cause in the preferential loss of radial arcs at a given image
surface brightness, we re-examined the selection of tangential arcs
after artificially increasing the background noise by a factor 3.  Of
the 38 tangential arcs (with $L/W>$7) observed in the Smith sample
(see Table~\ref{tab:arcsum}), 32 were still identifiable as arcs after
the background noise was increased.  This implies that $\sim$20\% more
radial arcs would be found if they were looked for at an identical
surface brightness limit as the tangential arc population.  On its
own, this systematic effect does not effect our conclusions on the
mean dark matter density profile in this cluster sample, as will be
shown in \S~\ref{sec:results}.

When searching for radial arcs, our strategy of examining images after
central galaxy subtraction is best suited for the case of a single,
dominant central galaxy.  However, eight clusters in our sample
contain multiple bright galaxies in their core for which our central
galaxy subtraction technique is less effective.  The conservative
results presented in \S~\ref{sec:results} do not change within the
uncertainties if these clusters are excluded from our study.  The
photometric properties and length-to-width ratio ($L/W$) of arcs found
in these clusters are also less certain than those found in clusters
dominated by a single central galaxy (see Table~\ref{fig:arctable} for
details).

\subsection{Arc Photometry \label{sec:arcphot}}

Photometric magnitudes were measured for all candidate arcs. This is a
complex task for two reasons.  First, arcs are by definition often
highly distorted, making them poorly-suited to automatic source
identification codes such as {\sc SExtractor} (Bertin \& Arnouts
1996). Second, contamination from bright, nearby galaxies can affect
the result, particularly for the radial arcs buried in galaxy halos.

\begin{figure*}
\begin{center}
\mbox{ \epsfysize=10cm \epsfbox{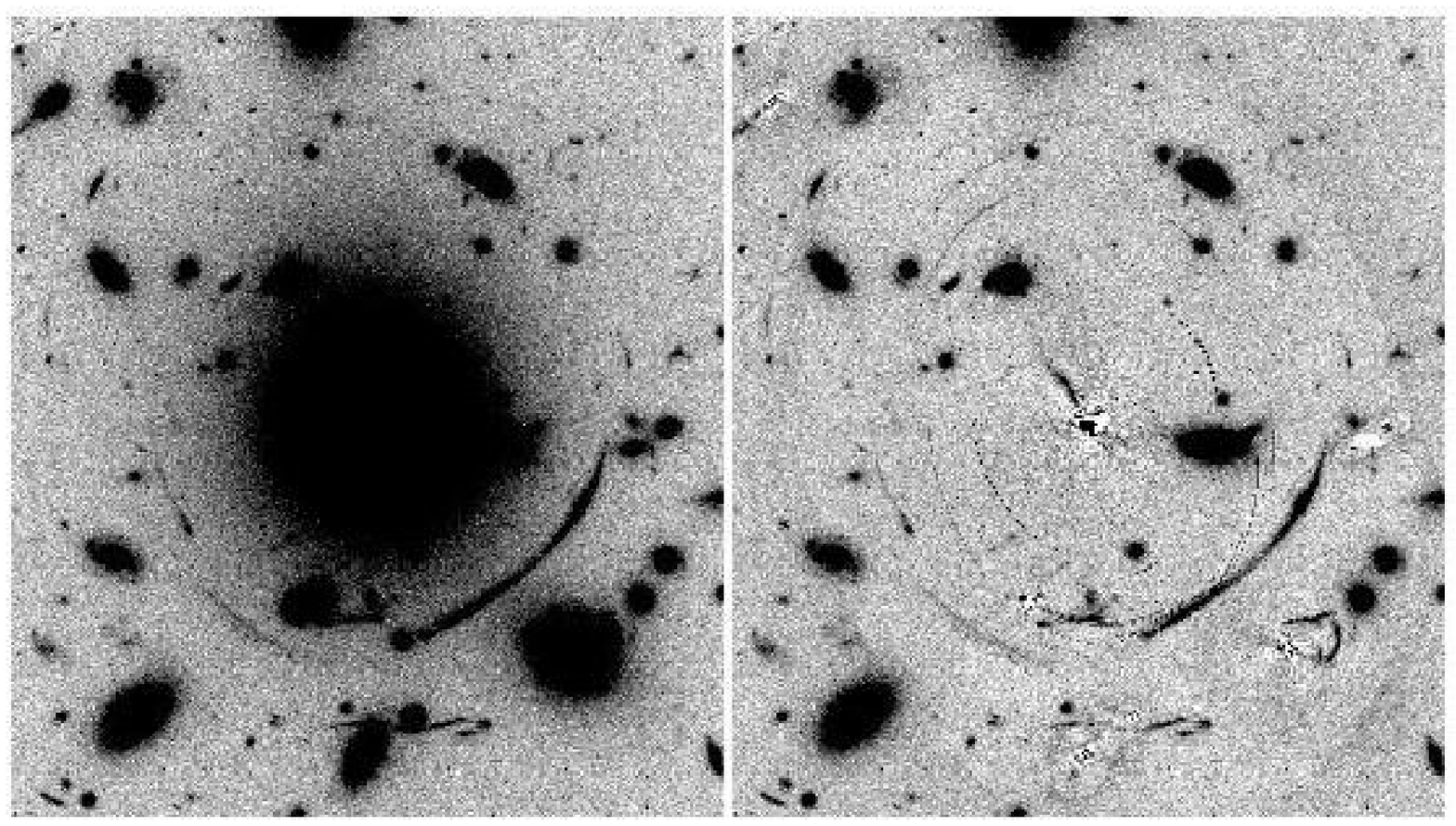}}

\caption{An example of galaxy subtraction performed to secure
photometry and length to width ratio for the arcs in Abell
383. Although the galaxy subtraction process leaves
tangentially-oriented residuals, these are easily distinguished from
true arc candidates by visual inspection. Radial arc candidates will
not generally be confused with these residuals.\label{fig:gsub}}
\end{center}
\end{figure*}

Our procedure was as follows. Polygonal apertures were determined for
each arc using the {\sc iraf} task {\sc polymark}.  Additionally, all
possible galaxy interlopers were digitally subtracted with {\sc
ellipse} (in addition to the BCG; see \S~\ref {sec:redux}), as
illustrated for Abell 383 in Figure~\ref{fig:gsub}. In order to
measure the photometric uncertainties, a master ``sky'' frame was made
by summing the initial sky subtraction of the image (see
\S~\ref{sec:redux}) and the subtracted contaminating
galaxies. Identical apertures were applied to both the ``sky'' frame
and the object frames to determine the magnitude and associated
uncertainty for each arc.

The subtraction of flux from cluster galaxies adjacent to an arc is
not a perfect process. The subtraction of these galaxies leaves
residuals which typically appear as thin ($<2-3$ pixels), tangentially
oriented features (see Figure~\ref{fig:gsub}). Fortunately, upon close
inspection these are readily distinguished from true gravitational
arcs.  Other residuals arise from nearby WFPC2 chip boundaries, tidal
features in the cluster, double nuclei in BCGs, dust-lanes, and spiral
arms. Those arc candidates whose photometry appears to have been
compromised due to such residuals are flagged in
Table~\ref{fig:arctable} in the Appendix.  Both the photometry and
measured $L/W$ for these flagged objects are more uncertain than the
formal uncertainty listed in the table.

\subsection{Arc Length-to-Width Ratio \label{sec:lwsec}}

The arc length-to-width ratio ($L/W$) is often used for characterizing
how strongly a source has been lensed. Limiting our arc sample
according to some $L/W$ criterion provides a means for undertaking
comparisons with earlier work and with theoretical predictions
(e.g. Bartelmann \& Weiss 1994). 

In practice, we measured $L/W$ ratios at three different signal to
noise per pixel thresholds: 2.0,1.5 and 1.0.  The mean and rms of
these three measurements is given in Table~\ref{fig:arctable}. All
$L/W$ measurements were done on the polygonal apertures used for
photometry, limiting the possibility of contamination from nearby
sources, and also limiting the chances of a very spurious $L/W$
measurement.  In the case where a cluster has multi-band data, the
final $L/W$ ratio is the mean found across the bands and the
uncertainty includes measures in all bands. Arc lengths are measured
by first finding the intensity weighted centroid of the arc within the
same polygonal aperture that was used to obtain photometry.  From
there the pixel furthest from the arc centroid above the threshold
$S/N$ is calculated.  Finally, the pixel (above the $S/N$ threshold)
furthest from this pixel is found.  The final arc length is the sum of
the two line segments connecting these three points.  The width is
simply the ratio of the contiguous area above the $S/N$ threshold to
the length.  If the arc width was found to be $<$0\farcs3 (the typical
WFPC2 PSF is $\sim$0\farcs15), then the feature was determined to be
unresolved in that direction.  In this case, the measured $L/W$ is a
lower limit (with the width set to 0\farcs3) as noted in
Table~\ref{fig:arctable}.  As discussed in the arc photometry
subsection, there are several arcs whose $L/W$ measurement was
possibly compromised due to residuals from the galaxy subtraction
technique, and these arcs have been flagged in
Table~\ref{fig:arctable}.

Since this is the first systematic search for radial arcs, all
candidates are presented in Table~\ref{fig:arctable} with an
accompanying finding chart in Fig~\ref{fig:radarcs}, regardless of
their $L/W$.  For the tangential arcs, only those with $L/W>7$ are
presented unless there is a spectroscopically-confirmed redshift in
the literature (even though these arcs were not used in our final
analysis, \S~\ref{sec:results}).

\subsection{Spectroscopic follow-up}

\begin{figure*}
\begin{center}
\mbox{
\mbox{\epsfysize=6cm \epsfbox{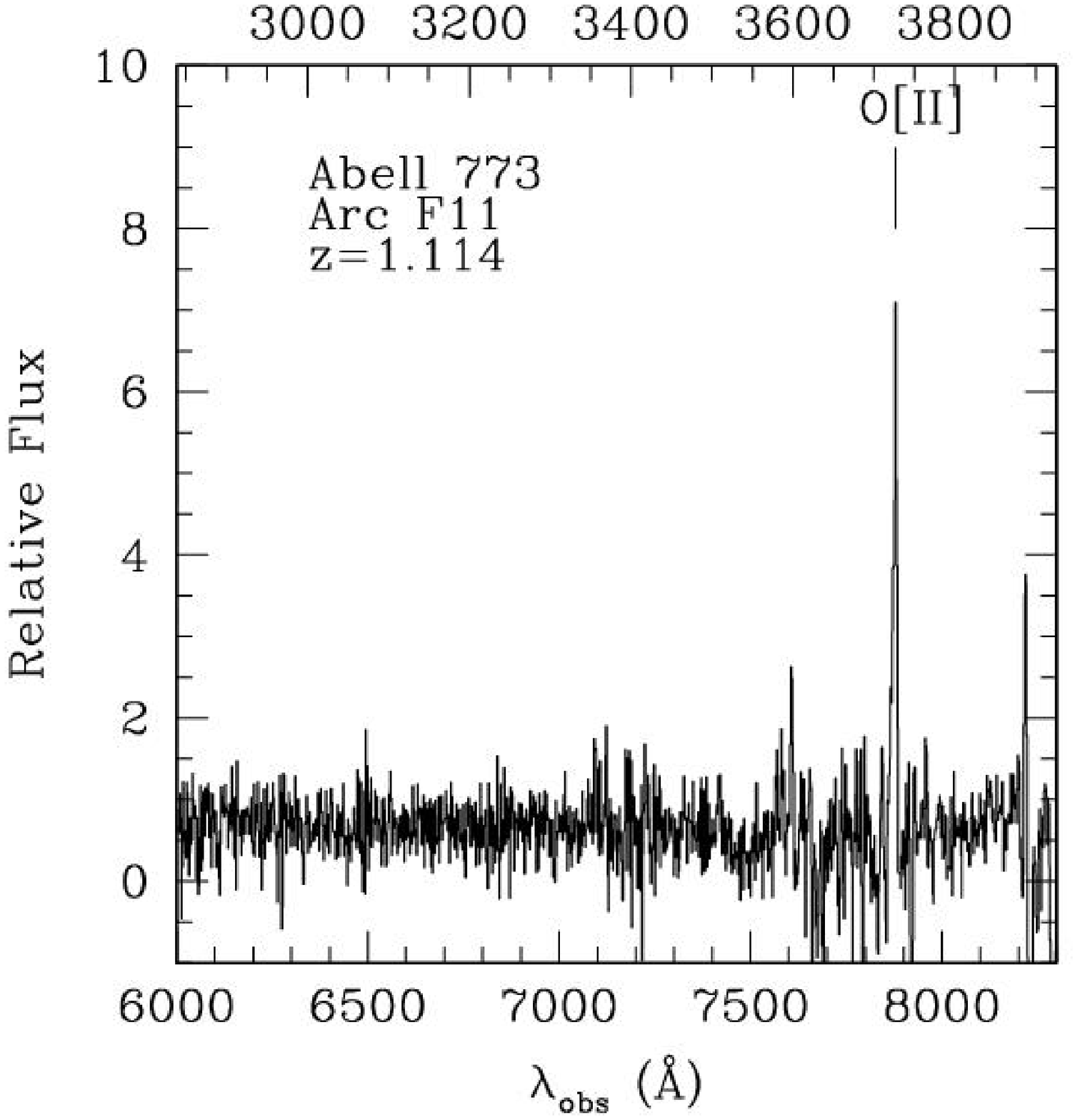}}
\mbox{\epsfysize=6cm \epsfbox{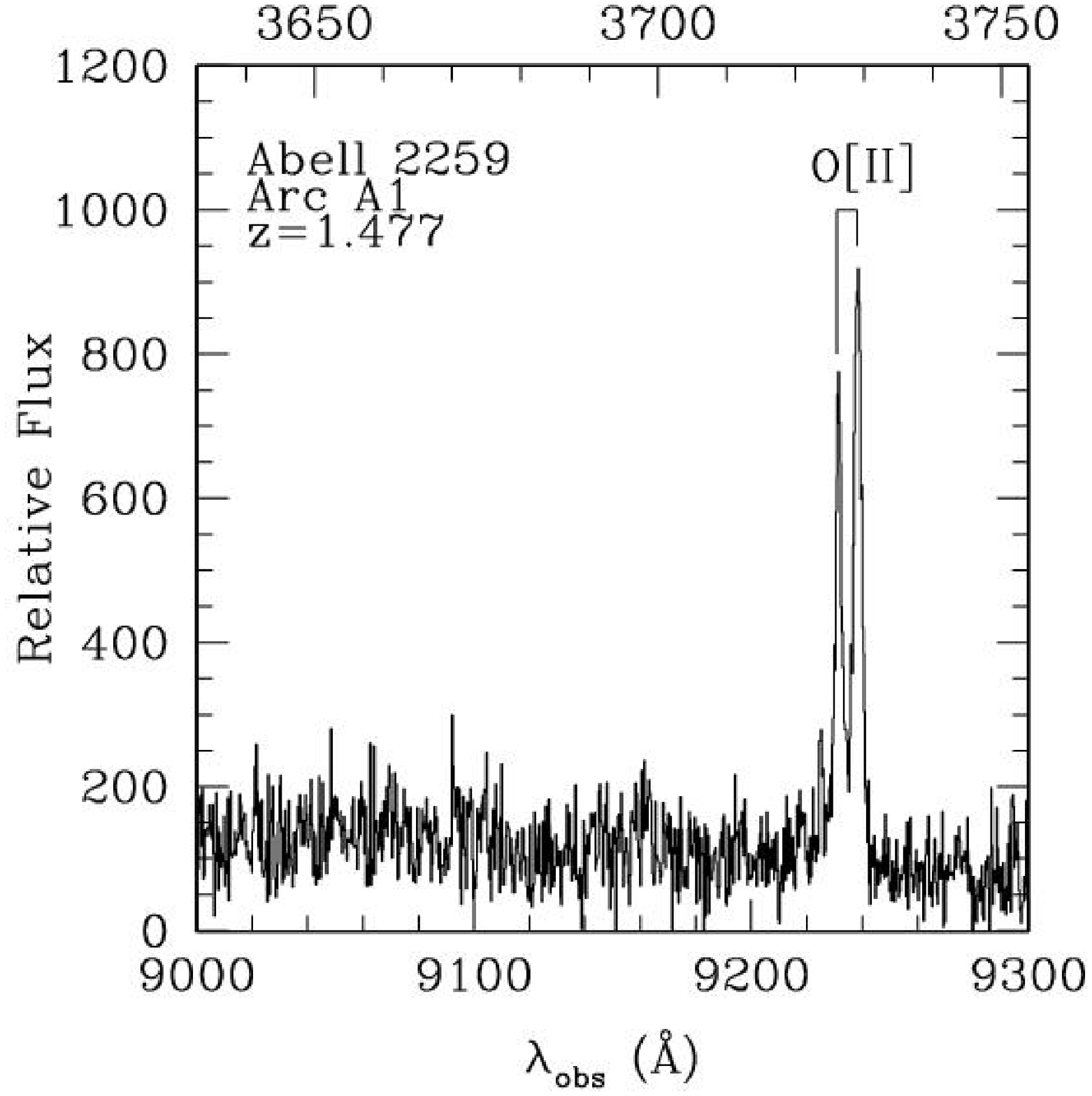}}
\mbox{\epsfysize=6cm \epsfbox{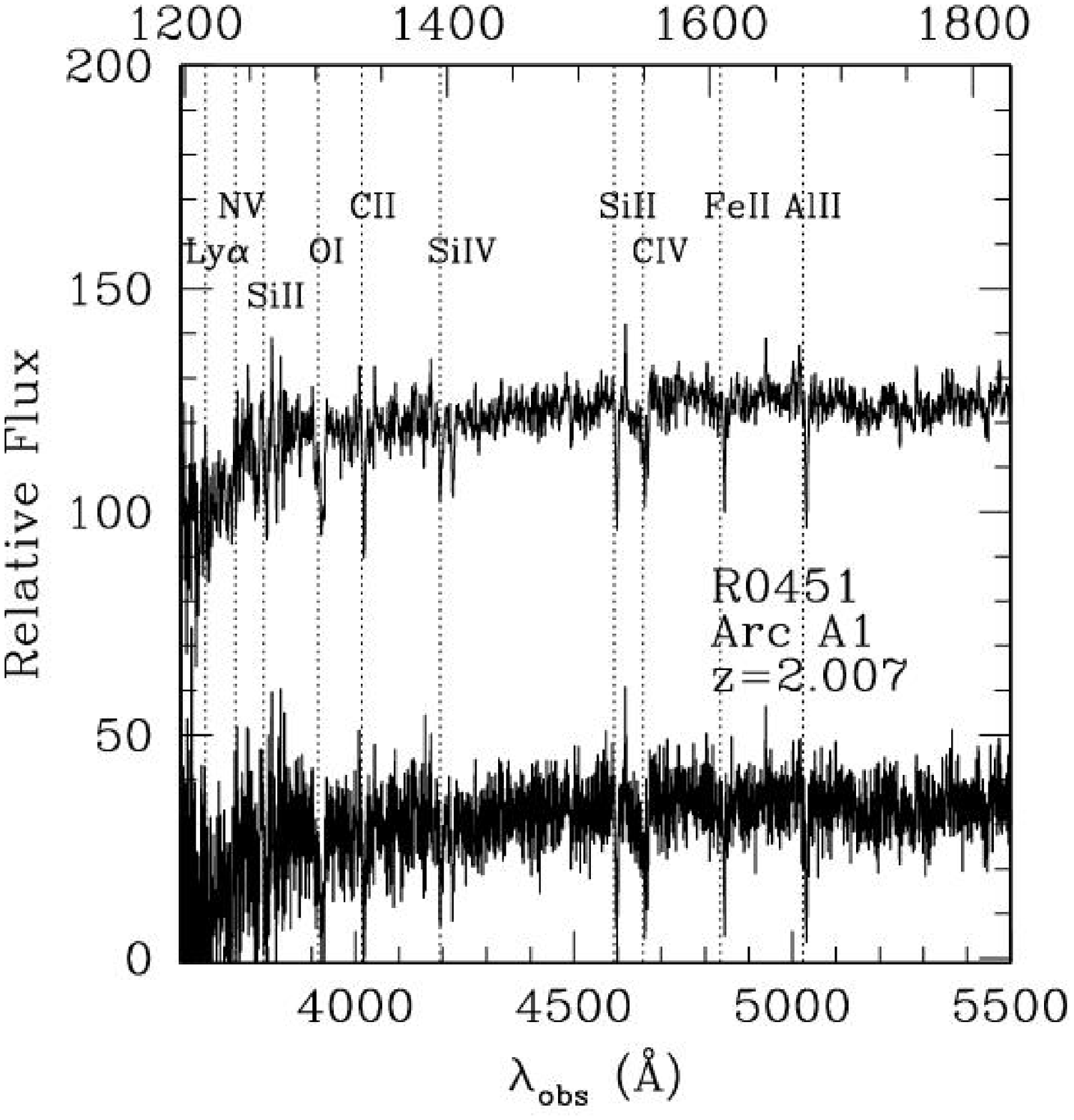}}
}
\mbox{
\mbox{\epsfysize=6cm \epsfbox{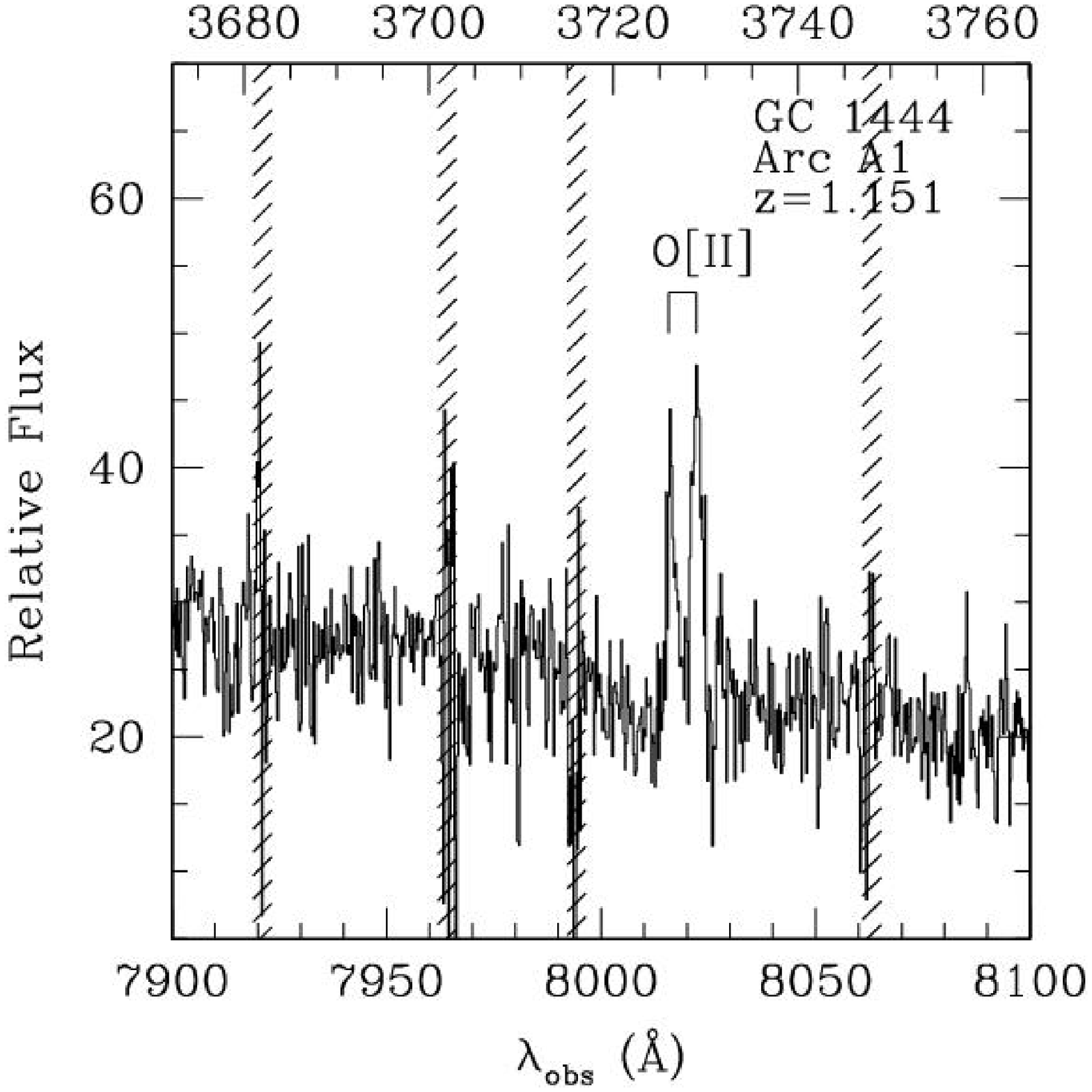}}
\mbox{\epsfysize=6cm \epsfbox{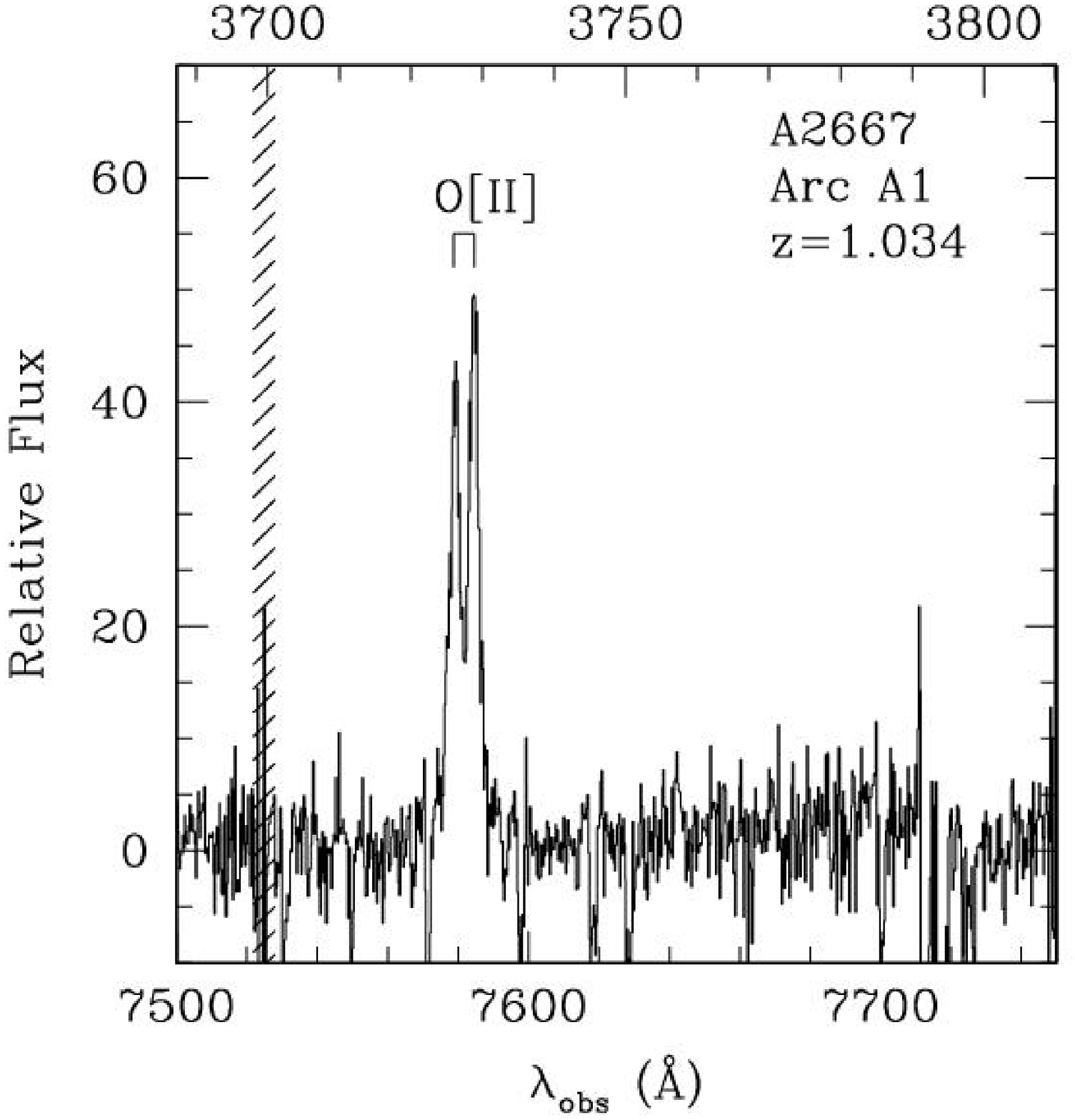}}
\mbox{\epsfysize=6cm \epsfbox{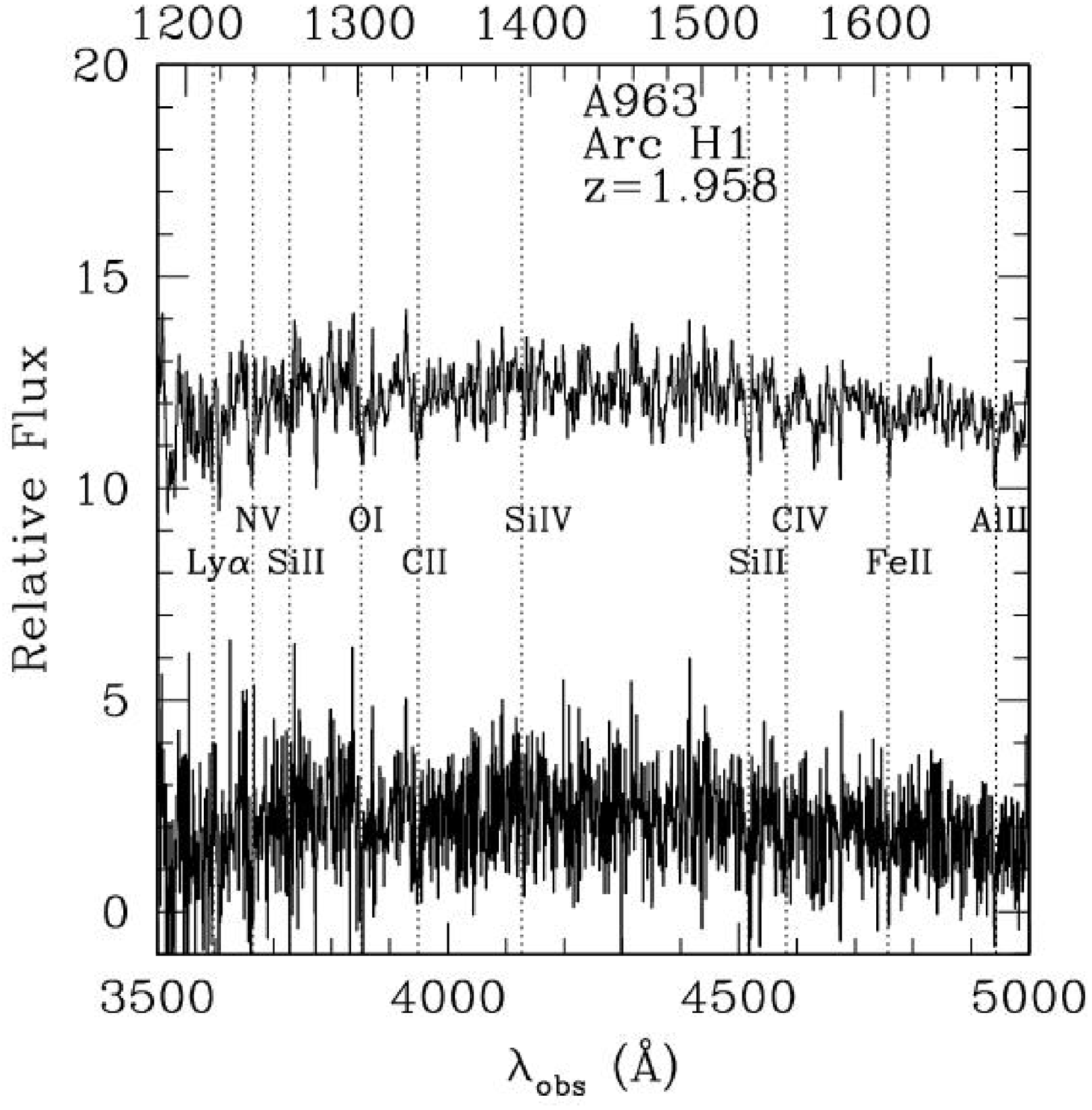}}
}
\mbox{
\mbox{\epsfysize=6cm \epsfbox{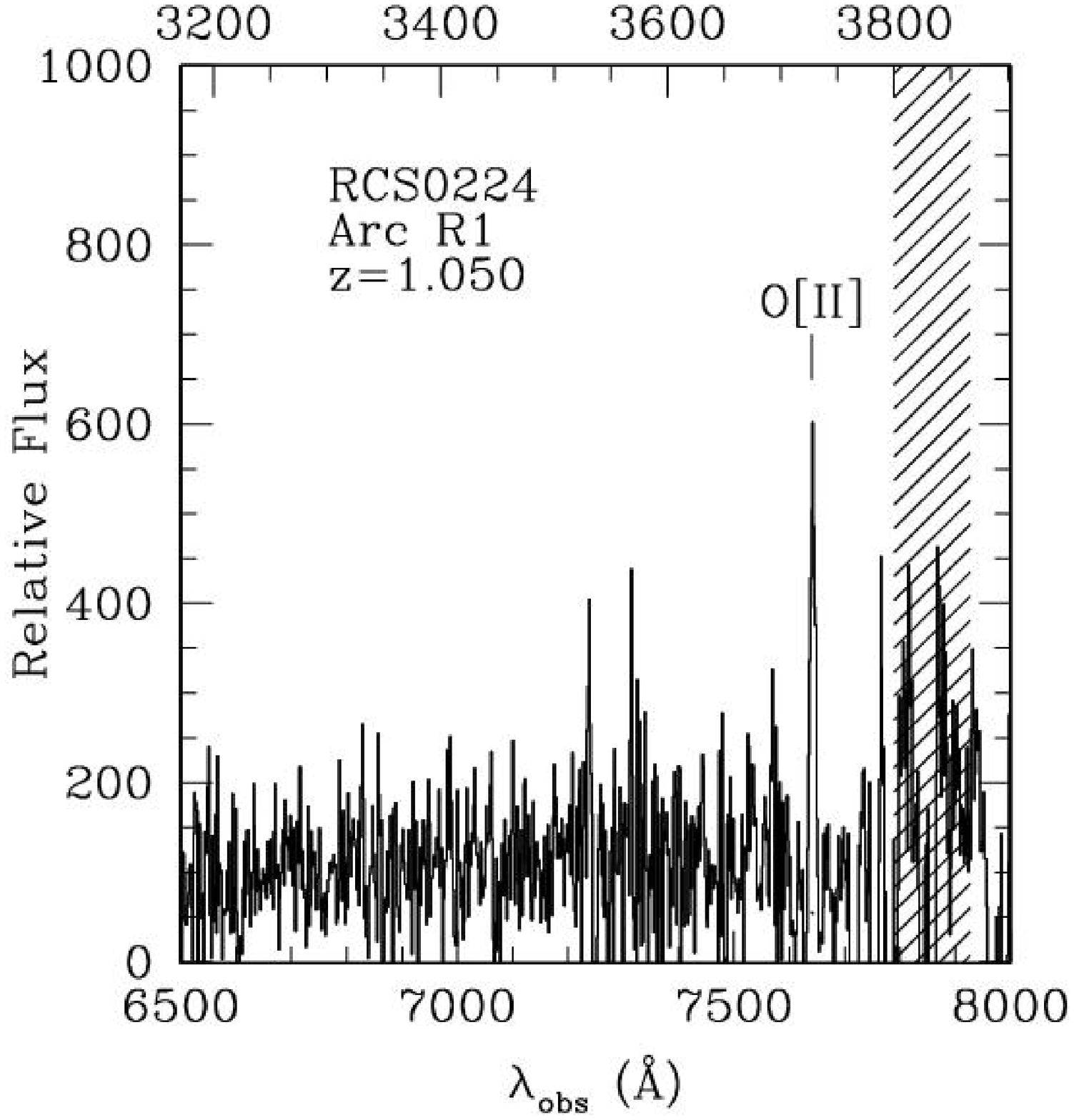}}
}
\caption{New gravitational arc redshift measurements.  The new radial
arc redshifts are in GC1444 and R0451.  A smoothed version of the arc
spectrums in Abell 963 and R0451 are also pesented so that the weak
absorption features can be more readily discerned.\label{fig:spectra}
}
\end{center}
\end{figure*}

To gauge possible contamination of the radial arc candidate list by
non-lensed sources, we have undertaken a limited Keck spectroscopic
campaign as part of our quest to obtain deep spectroscopy of lensed
systems for detailed individual study (Sand et al. 2002;
2004). Possible sources of contamination in the arc candidate list
include optical jets and cooling flow features associated with the
central cluster galaxy, and foreground edge-on disk galaxies.  A
summary of the spectroscopic results are given in
Table~\ref{tab:spec}. In this table we also present a compilation of
spectroscopic redshifts for several tangential arcs (those which have
not yet been published).  The new arc spectra are shown in
Figure~\ref{fig:spectra}.

Discussing the radial arc candidates in more detail:

\begin{itemize}

\item{} {\it GC 1444 \& RCS 0224:} These are the new radial arc
 redshifts presented in this work, based on single emission lines
 assumed to be O[II].  Both radial arc candidates have continuum
 blueward of the emission line making its interpretation as Ly$\alpha$
 unlikely.

\item {\it Abell 370, Abell 773, GC0848, Abell 1835, MS0440 and
AC118:} The spectra of these radial arc candidates were inconclusive.
The spectrum was either faint and featureless or not detected at all.
Note that spectra were taken for two radial arc candidates in AC118
(A1 and A2).  The spectral coverage of all observations was continuous
between $\sim$4000 and 10,000 \AA.

\item{} {\it MS 1455, 3c435a, 3c220, IRAS 0910 \& A2667:} These radial
arc candidates are sources at the cluster redshift. The spectra
exhibit numerous emission lines including: [OII], H$\gamma$, H$\beta$,
[OIII] 4959 and 5007, [O I] 6300 and 6363, [N II] 6548, H$\alpha$, [N
II] 6583, and [SII] 6716 and 6731. These features each have velocity
structures that can span hundreds of kilometers per second.

\end{itemize}

In addition to the spectroscopy presented in Sand et al. (2002, 2004),
we have now attempted spectroscopic verification for 17 candidate
radial arcs. Five spectra are consistent with the lensing hypothesis
(Abell 383, MS2137, GC1444, RCS0224 (arc R1) and RXJ1133), seven are
inconclusive (Abell 370, Abell 773, GC0848, Abell 1835, MS0440 and two
arcs in AC118), and five turn out to be spectroscopically coincident
with the cluster redshift (MS1455, 3c435a, 3c220, IRAS 0910 \& A2667).

Although not all radial candidates selected for Keck spectroscopy have
$L/W>$7, it is fair to assume this sample is representative of the
archive catalog list, since targets were selected on availability at
the telescope (e.g.~RA \& DEC) and not towards arcs with any specific
quality.  A key issue, however, in deriving a contamination fraction
is the question of the identity of those candidate radial arcs whose
nature we were unable to confirm. Most likely these are either optical
synchrotron jets associated with the BCG (which would have featureless
spectra) or 1$<z<$2 lensed systems with a weak absorption line
spectrum.  Based on our current spectroscopy, we estimate that at
least $\sim$30\% of the radial arc candidates are likely to be
non-lensed features.  If we assume that half of the inconclusive
spectra are also contaminants than the fraction would increase to
$\sim$50\%.  The basic conclusions of this paper regarding the mean
inner DM density slope are not sensitive to even a $\sim$50\% decrease
in the total number of radial arcs.

New redshifts were obtained for five tangential arc systems and are
summarized in Table~\ref{tab:spec}.  One comment is warranted
concerning the redshift of the southern arc in Abell 963.  After
considerable effort, an absorption line redshift ($z$=1.958) was
finally obtained for this low surface brightness feature (H1 in Smith
et al. 2005) using the blue arm of LRIS on Keck I. Another portion of
the southern arc (H2) seems to have a similar spectrum but of lower
$S/N$.  The redshift and brightness of arc A1 in R0451 are
interesting.  At $z=2.007$ and $F606W=20.24\pm0.03$ this object is of
similar brightness as the highly magnified Lyman break galaxy cB58.

\subsection{Arc Statistics: A Summary}

To summarize, using the archive of HST/WFPC2 galaxy clusters we
have visually identified candidate gravitationally lensed
features, performed appropriate photometry and measured the $L/W$
ratios.  The arc catalog is presented in the Appendix as
Table~\ref{fig:arctable}, along with published redshifts where
available (or if presented in this work).

As this is the first systematic search for {\it radial arcs}, we list
{\it all} such features found in Table~\ref{fig:arctable} and provide
finding charts in Figure~\ref{fig:radarcs}. The charts present the
original HST image and a galaxy-subtracted version.  In the case of
the {\it tangential arcs}, since our subsequent analysis (\S 4 and \S
5) will focus only on those with $L/W>$7 we only list those which
either have a measured redshift or $L/W>$7.  Candidate arcs
demonstrated spectroscopically to be foreground non-lensed sources are
not included. However, where the spectroscopy is inconclusive, the
candidates are retained. Offsets from the brightest cluster galaxy are
provided in Table~\ref{fig:arctable} to aid in their identification.

As discussed in \S 4, our statistical analysis will be based on both
radial and tangential arcs with $L/W>$7. In the total cluster sample
we have found 12 radial arc candidates and 104 tangential arc
candidates out of a total sample of 128 galaxy clusters. In
Table~\ref{tab:arcsum} we summarize the arc statistics for both the
total sample and those subsamples introduced in \S 2.1. The 68\%
confidence range for the radial to tangential number ratio was
computed using binomial statistics, appropriate for small number event
ratios (Gehrels 1986).  It is reassuring that the total sample and
subsamples give similar results for this ratio, since this implies that the
heterogeneous selection of clusters inherent in our analysis of the
archival data is unlikely to be a dominant uncertainty.

\section{Deriving Mass Distributions from Arc Statistics \label{sec:modelsec}}

In this section we discuss our methodology for calculating the
expected number ratio of radial to tangential arcs, which is based on
the precepts developed by O02 (\S 4.1).  We will include the effect of
the finite source size of the radial arc sources into our calculation,
confirming that this is a significant contributor to the radial arc
cross section which ultimately affects the deduced inner DM slope. We
introduce our mass model in \S 4.2 and show further that the effects
of a central BCG are also significant.  In \S 4.3 we summarize those
systematics that have been studied in previous analyses.  Finally, in
\S 4.4 we show how we use the arc cross sections to deduce the number
ratio of radial to tangential arcs.  Utilizing the tools presented in
this section, we will place constraints on the inner DM profile in
\S5.

\subsection{Methodology}\label{sec:method}

We follow the prescription presented by O02 for calculating the
expected number ratio of radial to tangential arcs. The lens
equation is given by (e.g.~Schneider, Ehlers, \& Falco 1992),

\begin{equation}
\label{eq:lenseqn} y = x - \alpha (x) = x - \frac {m(x)}{x},
\end{equation}

\noindent where $y$ and $x$ are scaled radii in the source and lens
plane, respectively.  Throughout this work, we choose the
generalized-NFW scale radius, $r_{sc}$, as our scaling radius, meaning
that $x=R/r_{sc}$ and $y=\eta D_{l}/(r_{sc}D_{s})$.  The deflection
angle, $\alpha$, is determined by the mass distribution of the lens
where the quantity $m(x)$ is defined by

\begin{equation}
\label{eq:promenc} m (x)=2  \int_0^{x}dy \kappa (y) y.
\end{equation}

\noindent The quantity $m(x)$ is proportional to the mass inside projected
radius $x$ and $\kappa (x)$ is the surface mass density scaled by
the critical surface mass density, $\Sigma_{cr}$,

\begin{equation}
\label{eq:kappa} \kappa (R) = \frac{\Sigma_{tot}
(R)}{\Sigma_{cr}},
\end{equation}

\noindent where

\begin{equation}
\label{eq:sdcrit} \Sigma_{cr} = \frac{c^{2}}{4 \pi
G}\frac{D_{s}}{D_{l} D_{ls}},
\end{equation}

\noindent and $D_{l}$, $D_{ls}$, and $D_{s}$ are the angular diameter
distance to the lens, between the lens and source and to the
source, respectively.

With these definitions, the two eigenvalues of the Jacobian
mapping between the source and image plane can be written as

\begin{equation}
 \lambda_{r}= 1-\frac{d}{dx}\frac{m(x)}{x},
\lambda_{t}=1-\frac{m(x)}{x^{2}}. \label{eq:eigen}
\end{equation}

\noindent The root of these two equations describes the radial and
tangential critical curves of the lens.  Since the magnification
of the source is equal to the inverse of the determinant of the
Jacobian, the radial and tangential critical curves define regions
where the magnification of the source formally diverges.  For a
simple spherical lens, an infinitesimal source at $x$ in the image
plane is stretched by a factor $\mu_{t}=1/\lambda_{t}$ in the
tangential direction and $\mu_{r}=1/\lambda_{r}$ in the radial
direction. For an infinitesimal source, the cross section for
either a radial or tangential arc is then simply the area in the
source plane where

\begin{equation}
\label{eq:rthresh} R(x) = \left|\frac
{\mu_{r}(x)}{\mu_{t}(x)}\right| \geq \epsilon
\end{equation}
\begin{equation}
\label{eq:tthresh} T(x) = \left|\frac
{\mu_{t}(x)}{\mu_{r}(x)}\right| \geq \epsilon,
\end{equation}

\noindent with $\epsilon$ being the minimum arc axis ratio to be
considered.  O02 demonstrates that only for $\epsilon\geq7$ is the
ratio of radial to tangential arcs relatively robust with respect to
systematic uncertainties such as source and lens ellipticity (see
\S~\ref{sec:sys}). Even then, the finite source size of radial arcs
must be taken into account. Throughout this work we will only consider
situations in which $\epsilon\geq7$, corresponding to $L/W \geq7$.

As in O02, we first take the source size to be small for typical
tangential arcs and so use Eqn.\ref{eq:tthresh} directly (see e.g.
Hattori, Watanabe, \& Yamashita 1997 for justification and Fig. 5;
left panel of O02).  Then the cross section for tangential arcs is

\begin{equation}
\label{eq:sigtan} \sigma_{tan}= \pi \left(\frac {r_{sc}
D_{S}}{D_{L}} \right)^{2} (max(|y_{t,+}|,|y_{t,-}|))^{2},
\end{equation}

\noindent where $y_{t,+}$ and $y_{t,-}$ correspond to the position on
either side of the tangential caustic which satisfies
Eqn.~\ref{eq:tthresh}. Figure~\ref{fig:Tfig} illustrates the
situation.

We now consider the effect of the finite source size for the radial
arc sources. Useful diagrams for illustrating the relevant geometry
are provided in Figs. 1 and 2 of O02 and we will adopt the
nomenclature and procedure of that work. We assume that the sources
are circular with a finite radius and consider situations where the
source touches, crosses or lies within the radial
caustic. Figure~\ref{fig:srcsz} illustrates the dramatic effect that
the finite source size can have on the radial arc cross section. 

To allow for this important effect we require the true (unlensed) size
distribution of a representative sample of $z\simeq$1-1.5 galaxies
typical of those being lensed by our clusters.  Fortunately, a
$z\sim$1.4 size distribution has been presented by Ferguson et
al. (2004) based on the GOODS survey (see Fig. 2; top panel, of that
work), and we will adopt this for the remainder of our
analysis. Galaxy sizes in this redshift bin were found to have
half-light radii (which we will take as the radius of our sources)
between $\sim$0\farcs2 and $\sim$1\farcs1 with the peak of the
distribution at $\sim$0\farcs7.  We shall show later that the arc
number ratio is relatively insensitive to source redshift, making this
redshift bin choice unimportant (although it does roughly correspond
to that observed for typical arcs in our sample).

\subsection{Mass Models \label{sec:massmod}}

For the density profiles of our clusters we will adopt a simple,
spherically symmetric two-component mass model.  The simplicity of
this model is justified by O02, who showed that the number ratio
is a relatively robust quantity with respect to ellipticities in
the cluster mass distribution (see \S~\ref{sec:sys}).

The adopted model comprises the DM halo of the galaxy cluster (as
represented by the gNFW profile) and a luminous baryonic
component, representing the central cluster galaxy.  Previous work
on constraining the inner DM slope $\beta$ through the number
ratio of arcs neglected the possibly important contribution of the
BCG luminous component or concluded that the effects are small
(Molikawa \& Hattori 2001). Given that most of the radial arcs
found in our sample are buried in the halos of a bright, centrally
located galaxy (or a compact group of galaxies) it seems
appropriate to revisit this assumption. For example, it has been
shown numerically and theoretically (Meneghetti, Bartelmann, \&
Moscardini 2003) that the central cluster galaxy can increase the
cross section for radial gravitational arcs significantly,
especially if the underlying DM halo slope is shallow.

\subsubsection{Dark Component}

The cluster DM halo is modeled as

\begin{equation}
\label{eq:gnfw} \rho_d(r)=\frac{\rho_{c}(z)
\delta_{c}}{(r/r_{sc})^{\beta}\left[1+(r/r_{sc})\right]^{3-\beta}},
\end{equation}

\noindent which represents a generalization of the
numerically-simulated CDM halos, where $\rho_{c}$ is the critical
density and $\delta_{c}$ is a scaling factor.  This density profile
asymptotes to $\rho \propto r^{-\beta}$ at $r \ll r_{sc}$ and $\rho
\propto r^{-3}$ at $r \gg r_{sc}$. For values of $\beta = 1, 1.5$, the
DM density profile is identical to that found by NFW and nearly
identical to that of Moore et al (1998), respectively. Using this
general form for the DM halo allows for comparison to earlier
numerical results, although the latest generation of DM halo
simulations indicates that the DM profile may not converge to a simple
asymptotic slope (e.g.~Power et al. 2004; Tasitiomi et
al. 2004). Basic lensing relations for the gNFW form have been
presented elsewhere (e.g. Wyithe, Turner \& Spergel 2001).

The profile of the DM halo is characterized further by a
concentration parameter, $c_{vir}$.  In this work we follow O02
and Oguri et al. (2001) in determining the critical parameters of
the mean DM halo for a given mass. Following Bullock et al. (2001)
in characterizing the median and scatter in concentration
parameters for a given mass, we use

\begin{equation}
\label{eq:cvir} c_{vir}=\frac{r_{vir}}{r_{sc}},
\end{equation}
\begin{equation}
\label{eq:cvir2} c_{vir}=(2-\beta)c_{-2},
\end{equation}

\noindent where
\begin{equation}
\label{eq:c-2} c_{-2}=\frac{8}{1+z_{lens}} \left(
\frac{M_{vir}}{10^{14} h^{-1} M_{\odot}} \right)^{-0.13} ,
\end{equation}

\noindent where the factor of $(2-\beta)$ generalizes the situation to
$\beta \neq 1$ (Keeton \& Madau 2001).  From this relation, it is
possible to calculate both $r_{sc}$ and $\delta_{c}$ for a typical
halo of a given mass and inner DM density slope.

As Bullock et al. and others have found, there is significant
scatter around the median value of the concentration parameter.
Taking Eqn.~\ref{eq:c-2} and using the Bullock et al.  1-$\sigma$
dispersion around the median value of the concentration parameter

\begin{equation}
\label{eq:deltac}\Delta (log c_{-2})= 0.18,
\end{equation}
we have investigated the effect of a varying value of $c_{vir}$
(Figure~\ref{fig:ceffect}).  As can be seen, the arc cross section
ratio can vary by an order of magnitude between low and high
concentration halos of the same mass and inner slope.  For this
reason, the dispersion in halo concentrations will be taken into
account when we present our results in \S 5.  It is important to
note that the cross section ratio across concentrations is
relatively constant as a function of source redshift (see
\S~\ref{sec:x2arc}).

\subsubsection{Luminous Component}

Nearly all of the radial arc candidates discovered in our HST
search were either buried in the halo of a central BCG or that of
a compact multi-galaxy core.  It thus seems reasonable to include
a luminous baryonic mass component in our model. We used a
Hernquist (1990) mass density profile

\begin{equation}
\label{eq:jaffe} \rho_H(r)=\frac{M_{L} r_{H}}{2 \pi r (r_{H} +
r)^{3}},
\end{equation}
with total mass $M_{L}$ and $R_{e}=1.8153r_{H}$.  The Hernquist
luminous density distribution is found to be a good representation
of actual BCGs (see e.g. Sand et al. 2004).  Throughout this work
we choose $R_{e}$=25 kpc which is a typical BCG effective radius
(e.g.~Gonzalez et al. 2004).

Figure~\ref{fig:bcgeffect} illustrates the significant effect that
adding a massive central galaxy can have on the arc ratio. A major
degeneracy can be seen. The expected cross section ratio is similar
for both a $\beta$=1.0 and $\beta$=1.5 DM halo if a $10^{13}
M_{\odot}$ BCG is inserted.  Note again that the arc cross section
ratio is a relatively constant function of the background redshift for
a given mass model.  We also experimented with different values for
the BCG effective radius (15$<R_{e}<$45 kpc) and found that the
resulting number ratio varies by a factor of $\sim$5. Clearly the more
precise the information on the mass of the BCG, the more useful will
be the constraints on the DM profile.

\subsection{Summary of Systematic Effects \label{sec:sys}}

Here we summarize the model assumptions which affect our subsequent
analysis. The most thorough earlier investigation was by O02.  Oguri
considered the effect of finite source size, lens ellipticity, and
mass dependence of the cluster lens.

The finite source size greatly affects the radial arc cross-section
and O02 introduced an analytic formalism to correct for this. This
analytic formalism reproduces well the expectation from numerical
simulations and is used throughout this work. The tangential arc cross
section, on the other hand, changes relatively little as a function of
finite source size, particularly for the $L/W>$7 condition considered
here (e.g. O02; Figure 5).  Accordingly, no correction was made.

Lens ellipticity primarily changes the {\it absolute} number of
arcs (see also Bartelmann et al. 1995).  For a minimum axis ratio
of $L/W$=7, the arc ratio changes only by a factor of order unity.
Likewise, while changing the mass of the galaxy cluster has large
consequences for the absolute number of arcs expected, the effect
on the ratio of radial to tangential arcs for our minimum axial
ratio is also small.

It is for these reasons that the arc ratio is an attractive
statistic. No prior knowledge of the cluster mass is necessary and
relatively heterogeneous samples (such as the current HST sample)
may be used to find the average density profile.

Keeton (2001) studied the effect of source ellipticity on the number
ratio of arcs.  The basic conclusion of this work was that the ratio
of radial to tangential arcs decreases with increasing source
ellipticity.  The size of the effect is a factor of order unity.  For
the purposes of the present paper where we are applying the arc number
ratio test on the first observational sample of radial arcs, we will
not consider the effects of source ellipticity.

\subsection{The arc number ratio \label{sec:x2arc}}

Thus far we have presented predictions in terms of the ratio of arc {\it
cross sections} and we now take the final step towards a comparison with
the observations by determining the number ratio. To do this, for a given
mass, inner DM slope, concentration parameter and redshift, it is
necessary to integrate the product of the cross section and the number
density of galaxies over some range in background (source)
redshift. 

Since the arc cross section ratio is largely independent of source
redshift (see Figs.~\ref{fig:ceffect} and \ref{fig:bcgeffect}), the
expected number ratio of radial to tangential arcs should be
well-represented by the ratio of their cross sections, for a given
halo model. Moreover, the ratio is also fairly insensitive to the mass
of the underlying galaxy cluster. Therefore, with a single reasonable
fiducial model of fixed mass, lens redshift and source redshift we can
obtain constraints on the DM inner slope for the average galaxy
cluster in our sample.

It is not clear whether magnification bias will be significant in our
survey.  Our visually-based search method is not likely to be
flux-limited; it is largely the persistence of a relatively high
contiguous surface brightness signal that is noticed by a human
searcher.  As lensing conserves surface brightness, our arc search
should not be unduly affected by magnification bias.  Miralda-Escude
(1993) discusses the possible magnification bias in arc searches and
notes these may be significant in data affected by ground-based
seeing.  Given the improved point spread function of HST, it seems
safe to conclude that our search for resolved arcs is effectively
surface brightness limited.

\section{Results \label{sec:results}}

In this section we compare the observed arc number ratio with
theoretical predictions based on the methodology presented in
\S~\ref{sec:modelsec} and derive the first statistical constraints
on the inner slope $\beta$ of the DM distribution. We discuss the 
remaining  sources of observational error and review the prospects for
reducing their effect.

\subsection{Constraints on the inner DM slope and the role of the BCG}

Taking the formalism presented in \S\ref{sec:modelsec}, we now
calculate the expected number ratio of radial to tangential arcs as a
function of the DM inner slope, $\beta$, for a two-component
mass model.  We assume a fiducial model representing the typical
galaxy cluster in our sample with $M_{DM}$=1$\times 10^{15} M_{\odot}$
at $z_{lens}$=0.2 and a background at $z_{source}$=1.4.  As discussed,
we used the background size distribution from Ferguson et al. in order
to calculate the radial arc cross section.  We re-evaluate the
concentration parameter $c_{vir}$, according to Eqs.~\ref{eq:cvir2}
and \ref{eq:c-2} at each value of $\beta$.

\begin{figure*}
\begin{center}
\epsscale{0.75}
\plotone{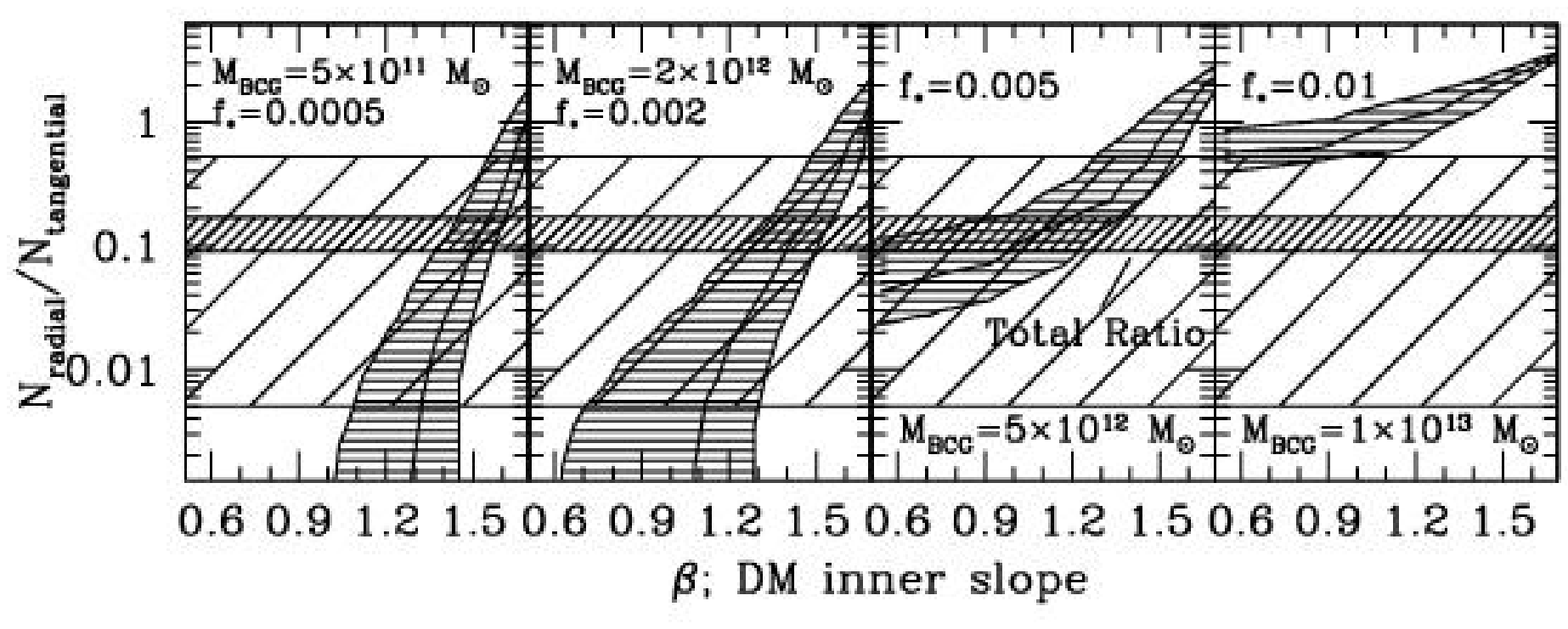}
\caption{Constraints on the inner DM profile, $\beta$, as a function
of the BCG mass (or BCG mass fraction, $f_{*}$, see text for details).
The narrow and wider hashed horizontal bars represent respectively
68\% confidence limits on the observed arc number ratio for the total
sample and the range in such limits for the individual cluster
sub-samples. The other band represents the range of theoretical
predictions if the concentration parameter is changed by $\pm1-\sigma$
for our fiducial cluster model. The left two panels span the BCG mass
range found in the detailed analysis of Sand et al (2004).  Virtually
no constraint on $\beta$ is found if the typical BCG mass is higher
than $5 \times 10^{12} M_{\odot}$ ($f_{*}=0.005$).
\label{fig:betaconstraint} }

\end{center}
\end{figure*}

\begin{figure*}
\begin{center}
\mbox{
\mbox{\epsfysize=4.0cm \epsfbox{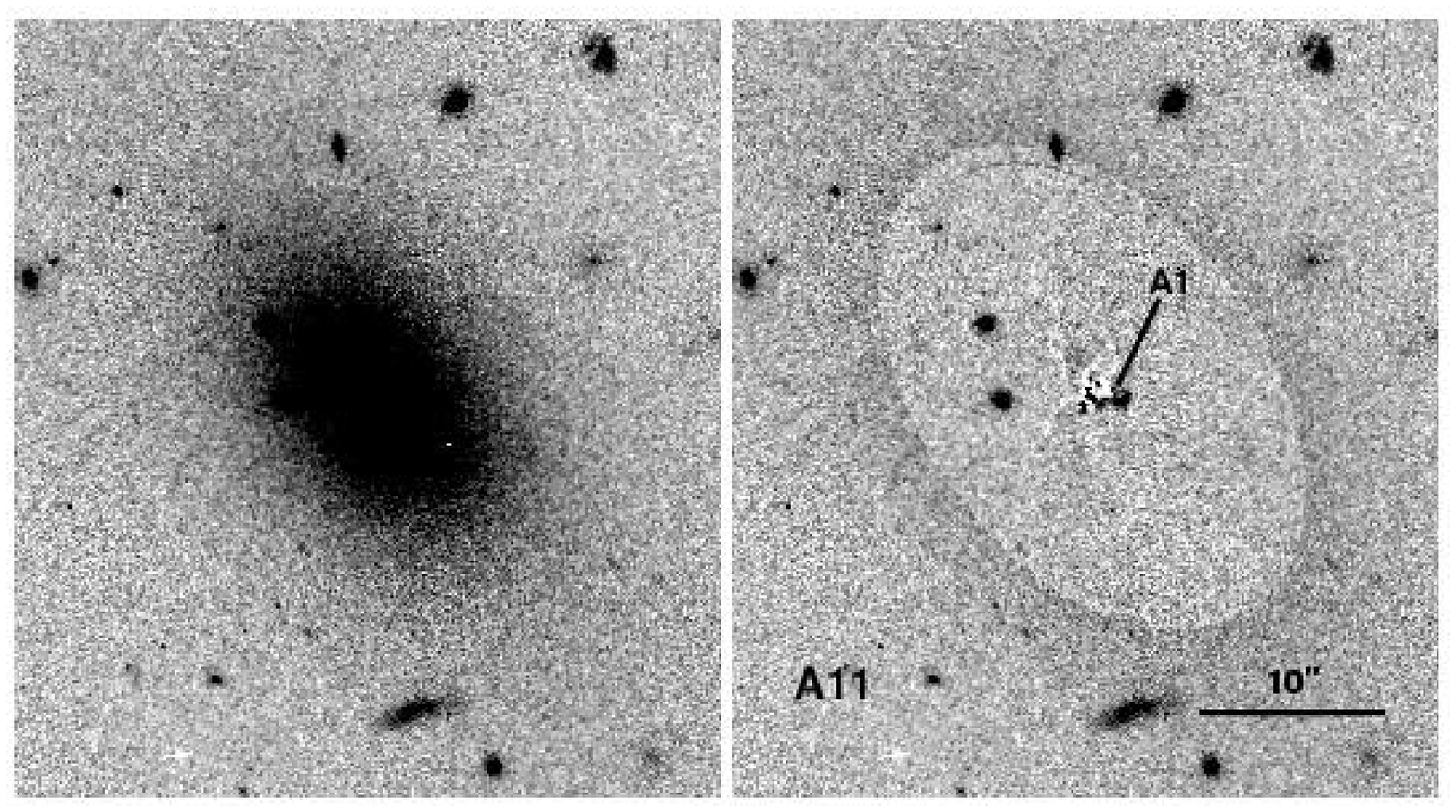}}
\mbox{\epsfysize=4.0cm \epsfbox{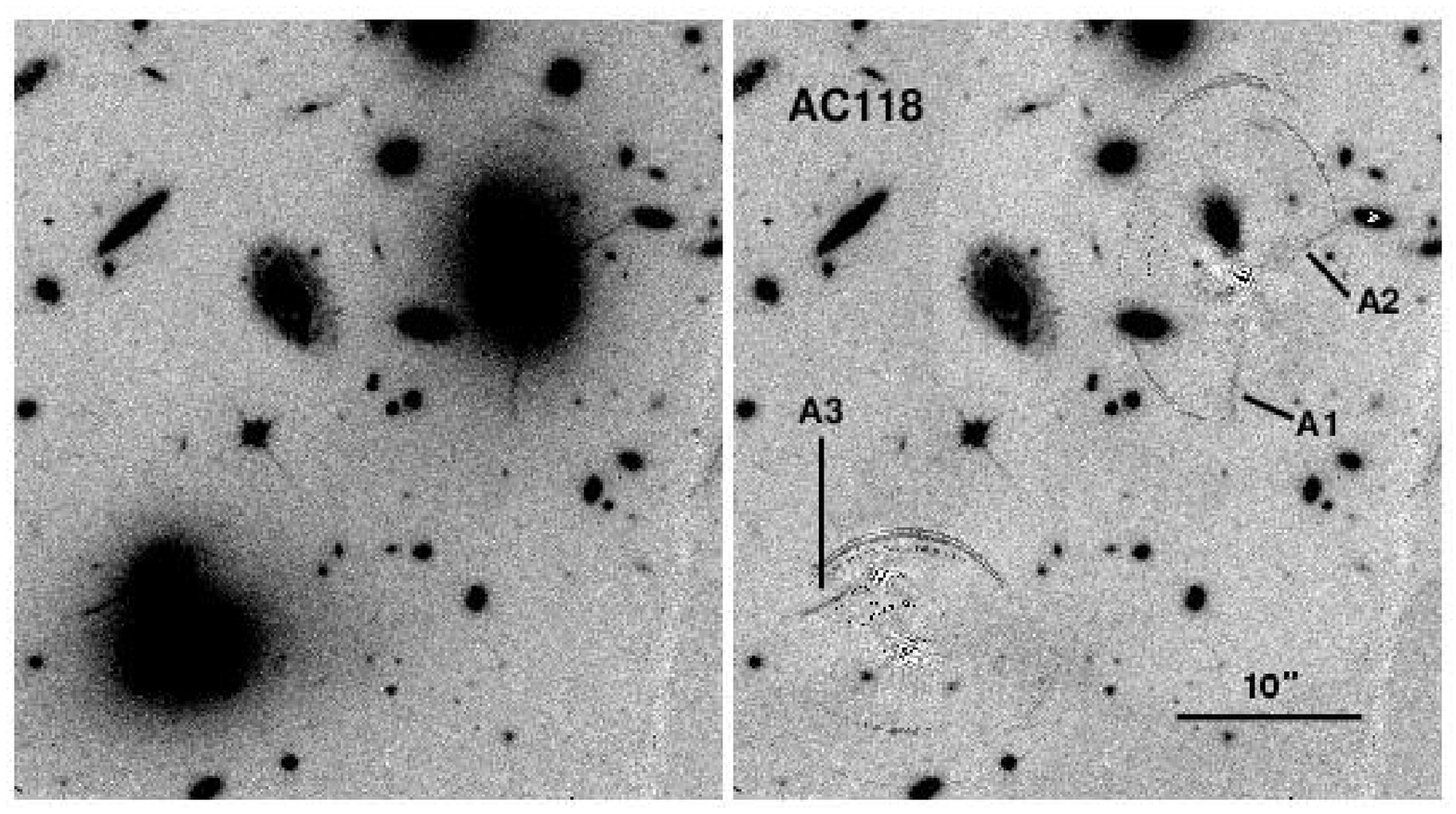}}
}
\mbox{
\mbox{\epsfysize=4.0cm \epsfbox{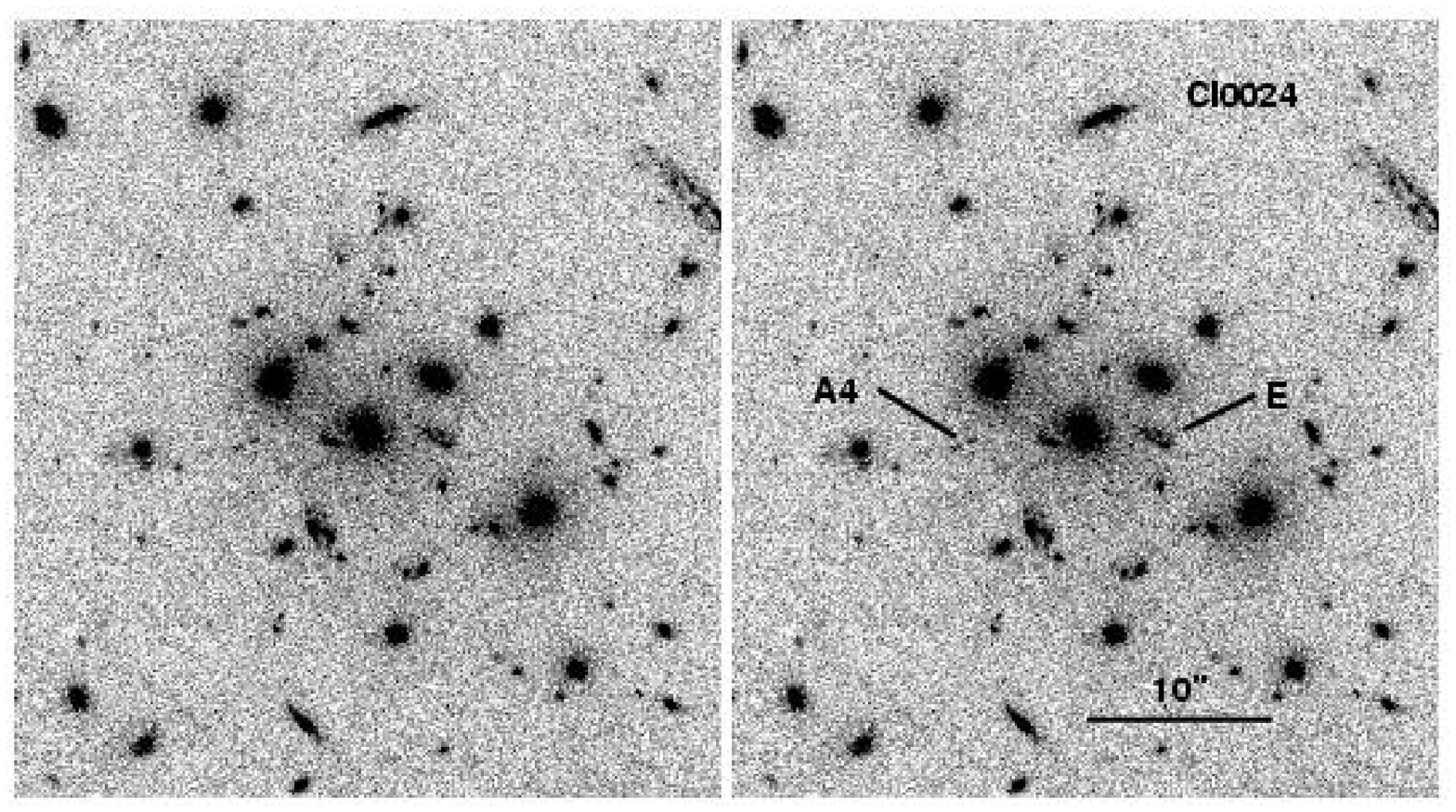}}
\mbox{\epsfysize=4.0cm \epsfbox{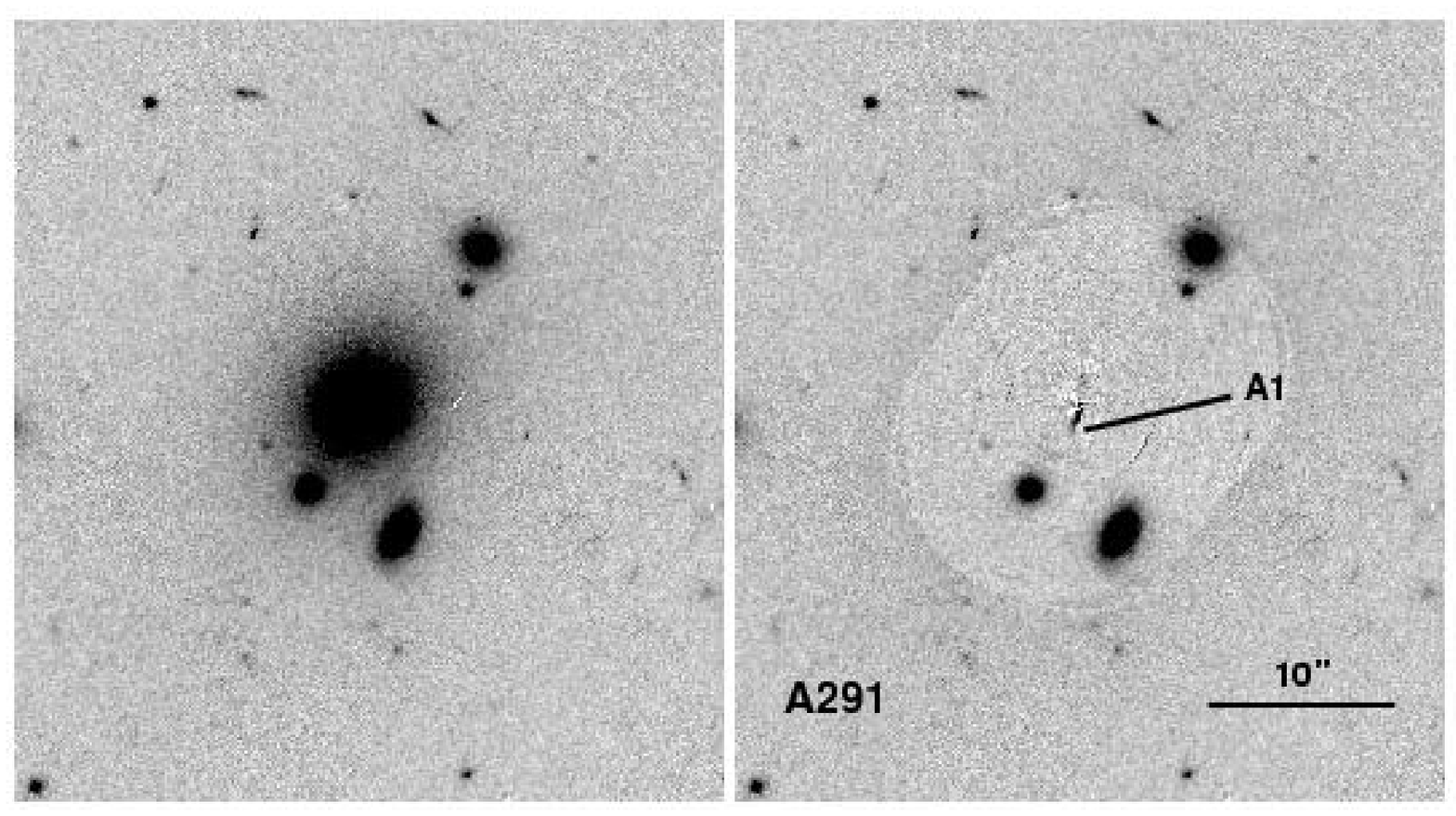}}
}
\mbox{
\mbox{\epsfysize=4.0cm \epsfbox{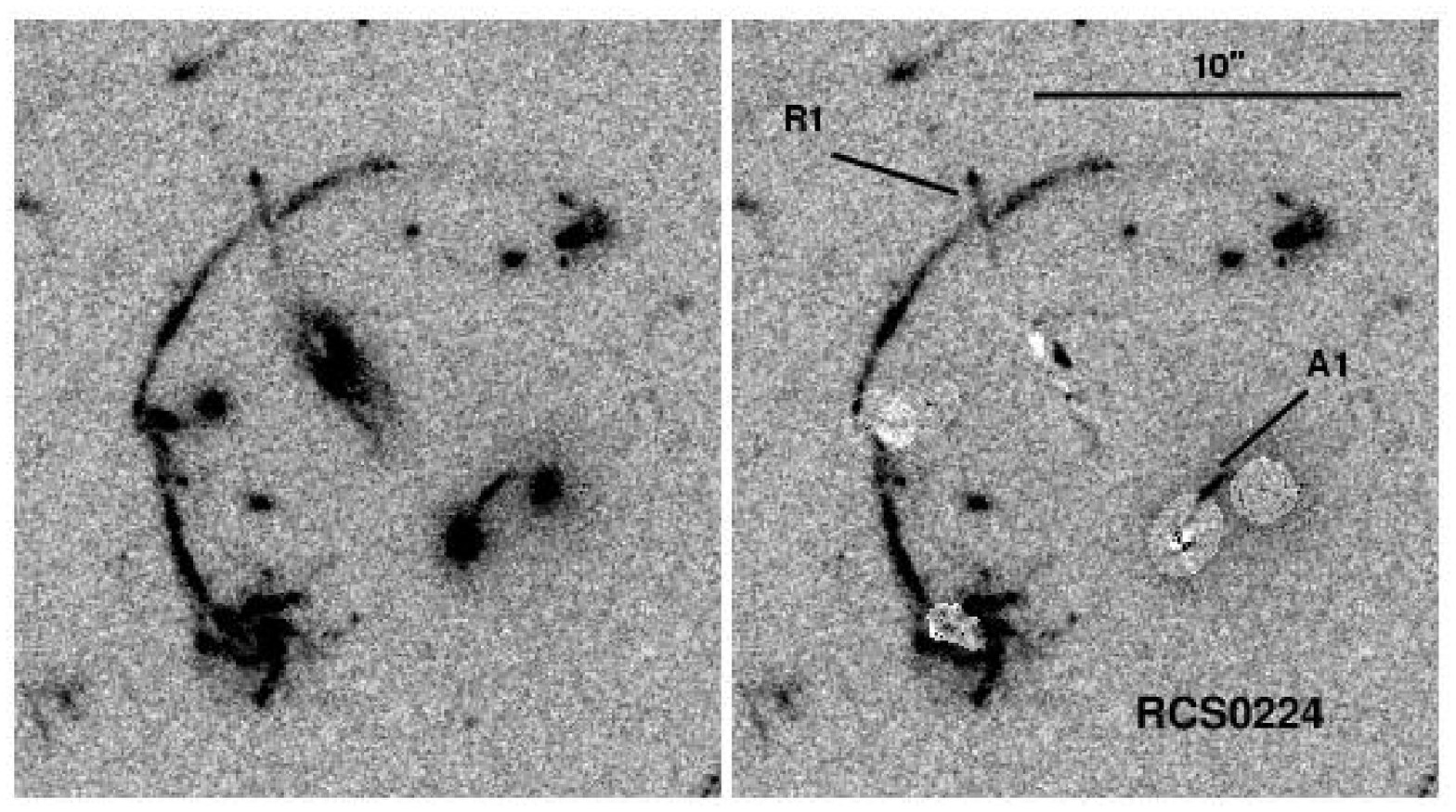}}
\mbox{\epsfysize=4.0cm \epsfbox{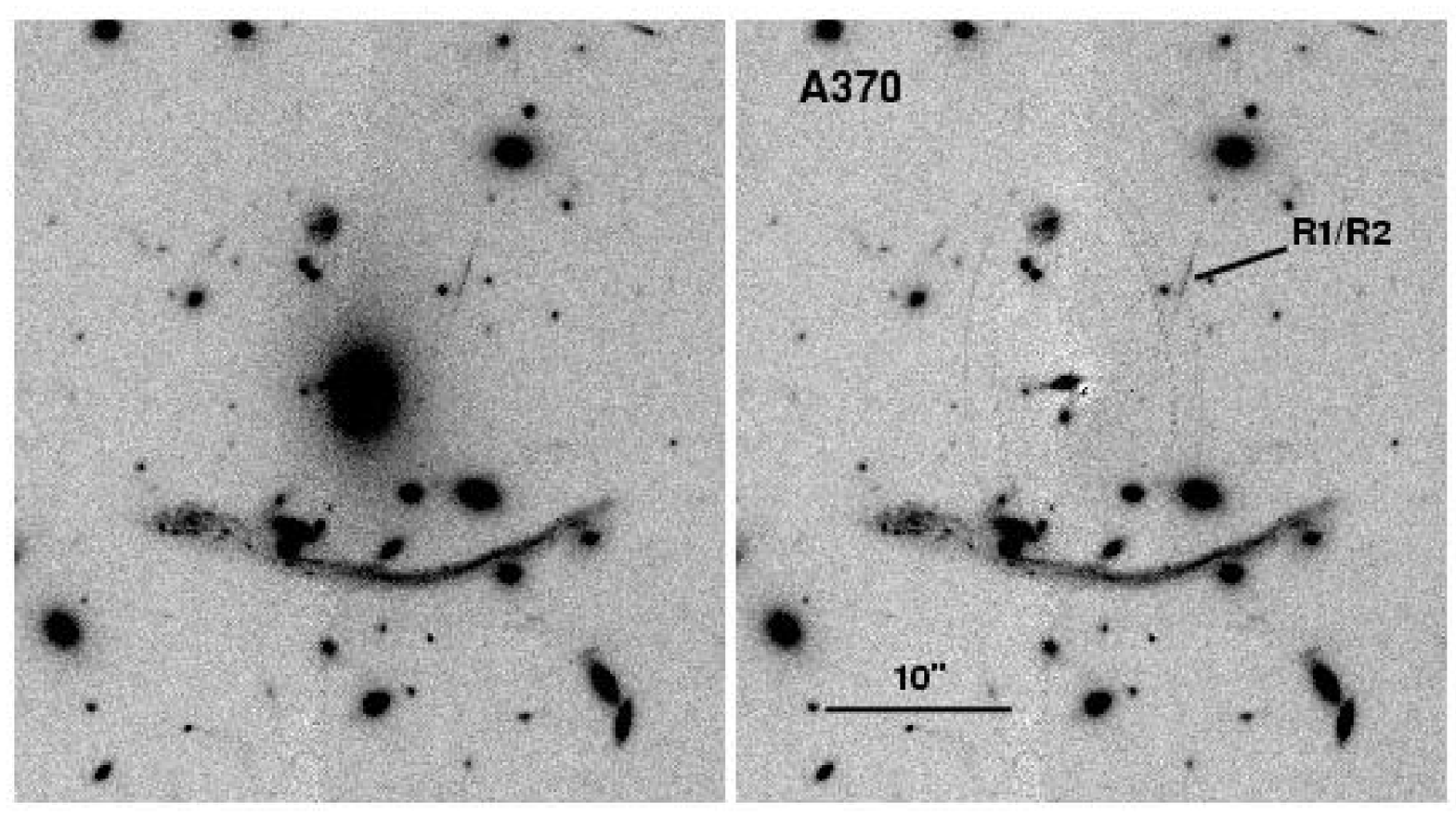}}
}
\mbox{
\mbox{\epsfysize=4.0cm \epsfbox{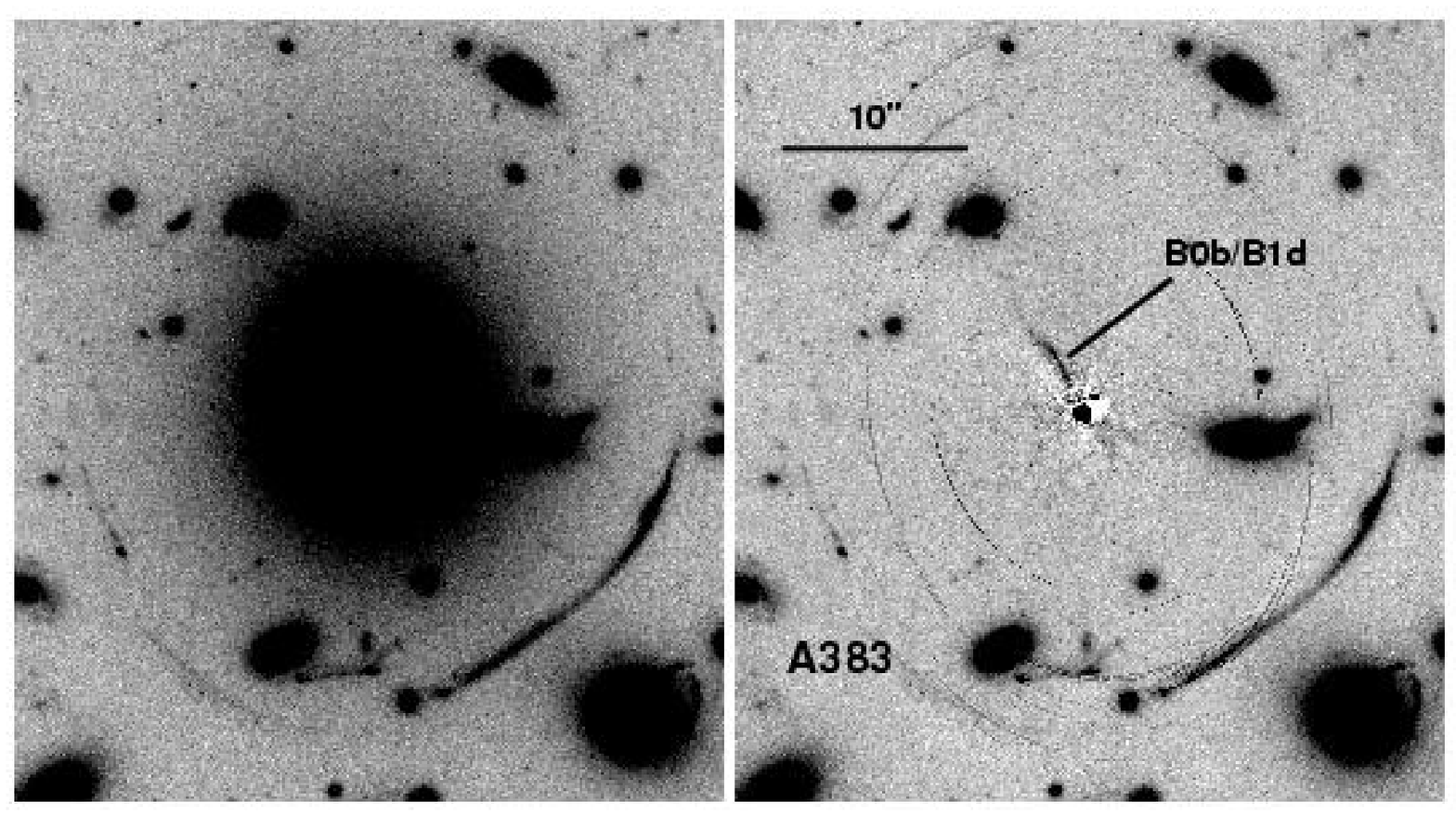}}
\mbox{\epsfysize=4.0cm \epsfbox{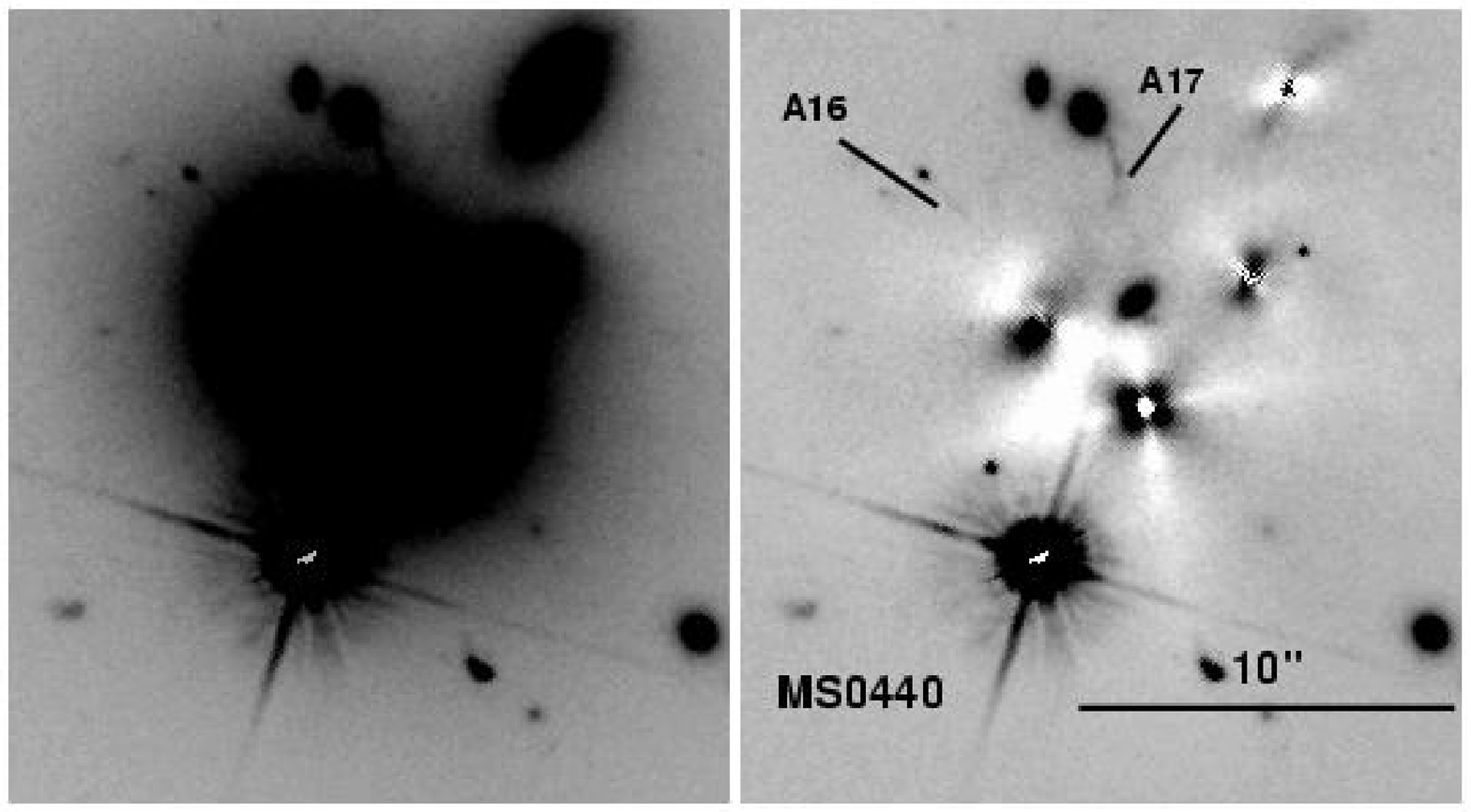}}
}
\caption{Radial arc finding charts \label{fig:radarcs}}
\end{center}
\end{figure*}

We present our constraints on the inner DM slope, $\beta$, in
Fig.~\ref{fig:betaconstraint}.  The two horizontal hatched bands
represent two estimates of the uncertainty in our measured radial to
tangential arc number ratio.  The inner, tightly hatched band
represents the 68\% confidence limit on the ratio across the total,
heterogenous, archival sample.  The outer horizontal band represents
the maximum range of the 68\% confidence regions amongst the cluster
subsamples presented in \S~\ref{sec:subsample} (see
Table~\ref{tab:arcsum}).  Although this enlarged region may not take
into account possible systematics associated with identifying arcs in
each sub-sample, it probably gives a reasonably cautious upper limit
on the uncertainties in the number ratio. The other, diagonally
oriented band (with horizontal hatches) in each panel shows the
predicted values of the number ratio for our fiducial cluster, taking
into account the expected 1-$\sigma$ dispersion of the concentration
parameter according to Eq.~\ref{eq:deltac}, given different BCG
masses.

The various panels in Fig.~\ref{fig:betaconstraint} represent
different assumed values for the mean stellar mass of the BCG,
recognizing that this is a key variable. The left two panels span the
range of stellar masses (5$\times$10$^{11}<M_{\odot}<$
2$\times$10$^{12}$) derived by quantitative dynamical and photometric
analysis in the sample studied by Sand et al (2004). The right-most
two panels represent more extreme stellar masses, the third
(5$\times$10$^{12} M_{\odot}$) being within the likely range, and the
fourth (1$\times$10$^{13} M_{\odot}$) somewhat extreme.  Also noted in
the panels of Fig.~\ref{fig:betaconstraint} is the BCG to DM mass
fraction, $f_{*}$, which is an alternative way to parameterize the
importance of the BCG for calculating the arc number ratio.

Depending on the mean BCG stellar mass, very different conclusions can
be drawn about the DM profile. If the Sand et al (2004) sample is
representative of the archive sample discussed here, the constraints
on the inner DM slope are reasonably tight with $1.2 \lesssim \beta
\lesssim 1.6$ for the total sample (and $0.7 \lesssim \beta \lesssim
1.7$ for the range spanned by the individual subsamples). In this
case, it would be reasonable to conclude the sample is consistent with
both NFW and the Moore profiles given the uncertainties. However, if
typical BCG masses are as high as 5$\times 10^{12} M_{\odot}$ we can
only constrain the dark matter density profile to have $\beta
\lesssim$1.3 (1.6 for the subsamples).  If the mean BCG mass were as
as high as $10^{13} M_{sol}$, then no acceptable solutions would be
found unless the true ratio of radial to tangential arcs was at the
upper end of the range of observed values.

Even if precise stellar masses for each BCG were available, it is
important to probe the sensitivity to the {\it cluster} properties.
We thus explored the effect on $\beta$ of changing the fiducial
cluster model, namely one with $M_{DM}$=1$\times 10^{15} M_{\odot}$ at
$z_{lens}$=0.2. In fact, reducing the cluster mass to
$M_{DM}$=5$\times 10^{14} M_{\odot}$ or changing the redshift to
$z_{lens}$=0.5 produces only a marginal change in the acceptable
values of $\beta$, illustrating again that the arc number ratio is a
robust tool if the BCG parameters can be constrained.

\subsection{Additional uncertainties and sample selection effects \label{sec:discuss}}

Of the uncertainties discussed earlier, those relating to the
identification and characterization of the {\it radial arcs} through
galaxy subtraction would lead to an {\it underestimate} of their true
number (and hence the radial to tangential number ratio), while
contamination from non-lensed radially oriented objects would work in
the opposite direction.  Remarkably, the total number of radial arcs
would need to change by nearly an order of magnitude for our
conclusions on $\beta$ to be significantly altered. This, we believe,
is highly unlikely given the tests we have performed.

For the {\it tangential arcs}, contamination may arise from chance
alignment of elongated foreground objects or tidal debris associated
with galaxies merging with the BCG.  Spectroscopic identification of a
large sample of arcs would be necessary to understand this
contamination rate in detail.  However, none of the tangential arcs
identified in \emph{HST} imaging and targeted in our Keck spectoscopic
program have turned out to be spurious. It seems safe to conclude that
the contamination rate is very low.  As mentioned in
\S~\ref{sec:arcid}, the tangential arcs found independently by two of
the authors disagreed only at the $\sim$10\% level providing an
estimate of the noise associated with visual identifications.

A further uncertainty related to our mass modeling technique is that
arising from cluster substructure.  In common with previous studies,
the modeling framework presented in \S\ref{sec:modelsec} assumes that
clusters comprise a single central DM halo spatially coincident with
the BCG.  However, Smith et al.\ (2005) show that 70\% of X--ray
luminous cores in their sample (Table~3) are unrelaxed with
${\sim}20$--60\% of the mass in structures not spatially coincident
with the BCG.

We explore the implications of possible substructure using the Smith
sample.  Interestingly, the radial to tangential arc ratio in the
Smith sample is lower than for the other cluster samples which may
arise from the substructure issues noted above.  To improve the
statistics, we also considered the larger Edge sample which was
selected to be dominated by relaxed systems (Edge, priv.\ comm.\, see
\S~\ref{sec:subsample}).  Contrasting the Edge (predominantly relaxed)
and Smith (predominantly unrelaxed) samples, we find the ratio
$r_{EDGE}$= $0.27^{+0.26}_{-0.14}$ and $r_{SMITH}$=
$0.03^{+0.06}_{-0.025}$, differing at the ${\sim}2$--${\sigma}$ level.

Conceivably, $r_{SMITH}$ is depressed relative to $r_{EDGE}$ by the
substructure present in the Smith clusters.  This idea is supported by
numerical simulations (Jing 2001) which show that DM halos in equilibrium
(relaxed) have higher typical concentrations than DM halos out of
equilibrium (unrelaxed).  If true, this would naturally explain the
different arc ratios seen in the Smith and Edge samples, since higher
concentration halos yield higher arc number ratios (see
Fig~\ref{fig:ceffect}).  In this respect, the scatter between the
subsamples, as indicated in Fig.~\ref{fig:betaconstraint}, may be a
reasonable measure of the effects of substructure.

This can also be understood in terms of lensing cross-sections (see
similar argument in Molikawa \& Hattori 2001).  Introducing
irregularities (substructure) into a cluster mass distribution
generally increases the shear, $\gamma$.  Since the tangential
critical line forms where $1-\kappa-\gamma=0$, increasing $\gamma$
pushes the tangential arcs towards regions with lower $\kappa$,
i.e. further from the center of the cluster, and thus the
cross-section to tangential arc formation increases.  The effect works
in the opposite sense for radial arcs, which form where
$1-\kappa+\gamma=0$.  Additional shear shifts radial arcs toward
regions with higher $\kappa$, i.e.\ closer to the center of the
cluster and thus reduces the cross-section to radial arc formation.
Therefore clusters with significant substructure are expected to
display a lower arc number ratio than clusters that are more
axisymmetric with little substructure.

It will be important to quantify this effect more rigorously in future
experiments that combine BCG mass estimates with arc number ratio
measurements.  A key aspect of such work would be to study a large,
well--defined sample of clusters (of order ${\sim}100$) for which both
homogeneous \emph{HST} data and reliable cluster substructure diagnostics
are available.

\section{Summary and Prospects}

In this work we have undertaken a systematic search for
gravitational arcs in the HST/WFPC2 cluster archive.  Since we
digitally subtracted bright cluster galaxies, this is the first
arc survey which is sensitive to radial gravitational arcs.  Using
this unique data set, we have attempted to place constraints on
the inner DM density slope, $\beta$, for this sample of clusters.

The main results of the paper can be summarized as follows:

1. A careful search of the 128 galaxy clusters reveals 12 radial
and 104 tangential arc candidates with a length-width ratio $L/W>$7.

2. Taking the entire sample of galaxy clusters, we have
constructed 3 smaller subsamples: two based on X-ray properties
and one with a roughly uniform optical imaging depth.  The arc
number ratio is roughly consistent across all three samples and
confirms the hypothesis that the radial to tangential arc number
ratio is a relatively robust statistic with respect to intrinsic
cluster properties.

3. Employing an analysis similar to that of O02, but with the
important addition of a BCG mass component, we have found that the
observed arc number ratio is consistent with a wide range of DM inner
slopes ($\beta<1.6$), depending on the assumed BCG mass.

The archive sample presented in this paper has illustrated a
potentially powerful method of constraining the profile of dark matter
in clusters. Although statistical in nature, as with all gravitational
lensing techniques, some assumptions are necessary. We have argued
that the {\it ratio} of the abundance of radial and tangential arcs
minimizes many of these leaving the mass of the baryon-dominated BCG,
cluster substructure and sample uniformity as the key issues. All of
these are, in principle, tractable problems given sufficient data.  We
thus remain optimistic that a valuable constraint on the distribution
of DM slopes can be derived via the methods described in this paper
given adequate observational effort.

\acknowledgements

A special acknowledgement goes to the WFPC2 instrument team and those
who maintain the HST archive.  Without their hard work this project
would not have been possible.  We would also like to thank the
anonymous referee for several clarifying comments.  We acknowledge
financial support from proposal number HST-AR-09527 provided by NASA
through a grant from STScI, which is operated by AURA, under NASA
contract NAS5-26555.  DJS would like to acknowledge financial support
from NASA's Graduate Student Research Program, under NASA grant
NAGT-50449.  TT acknowledges support from NASA through Hubble
Fellowship grant HF-01167.01. We thank Aaron Barth, Chuck Steidel and
Stanimir Metchev for helpful advice.  Finally, the authors wish to
recognize and acknowledge the cultural role and reverence that the
summit of Mauna Kea has always had within the indigenous Hawaiian
community.  We are most fortunate to have the opportunity to conduct
observations from this mountain.  This research has made use of the
NASA/IPAC Extragalactic Database (NED) which is operated by the Jet
Propulsion Laboratory, California Institute of Technology, under
contract with the National Aeronautics and Space Administration.

\appendix

\section{The cluster catalog}

Here we present our HST/WFPC2 cluster catalog used to identify
gravitational arcs.  The criteria for being included in the sample are
detailed in \S~\ref{sec:clussec}.  For each cluster entry, the
redshift, RA \& DEC, exposure time, filter, and X-ray luminosity are
listed.  If the cluster is associated with one of the three cluster
subsamples present in \S~\ref{sec:subsample}, this is noted as well.
A cluster is flagged if it has an associated arc which is presented in
Table~\ref{fig:arctable}.  The cluster RA \& DEC are taken from the
HST world coordinate system directly at approximately the position of
the BCG.

\section{The arc catalog}

Here we present the gravitational arc catalog derived from
Table~\ref{fig:clustable}.  For each arc, the redshift (if available),
magnitude, filter, length to width ratio, and offset from the BCG are
presented.  Also noted is whether or not the arc is radial or
tangential.  For the tangential arcs, only those with a $L/W>$7 are
listed, unless they have a spectroscopic redshift.  For the radial
arcs, all candidates are listed without regard to their $L/W$ value.
Note that only those arcs with a $L/W>$7 are included in the analysis
presented in \S~\ref{sec:results}.  If possible, the arc nomenclature
from previous work has been adopted.  Otherwise, an arc is labeled with
the prefix 'A' followed by a sequential number.  Those arcs whose
photometry and measured $L/W$ have possibly been affected by poor
galaxy subtraction are flagged (see discussion in \S~\ref{sec:arcphot}
\& \ref{sec:lwsec}).

\section{Radial Arc Finding Charts}

Here we present finding charts for all clusters with candidate radial
arcs, whether or not the arc has a $L/W>$7.  For each chart, the left
panel is of the original image, while the right panel is the
BCG-subtracted image from which the radial arc was identified, along
with a label corresponding to that presented in
Table~\ref{fig:arctable}.  Note that for Cl0024 and A1689, no bright
galaxies were removed in the finding chart.  Two finding charts are
presented for MS0451 since one of the radial arcs (A6) is associated
with a bright elliptical away from the cluster center.  No finding
charts were made of tangential arc candidates, although offsets from
the BCG are listed in Table~\ref{fig:arctable}.  North is up and east
is always towards the left hand side of the page.


\clearpage

\begin{inlinefigure}
\begin{center}
\resizebox{\textwidth}{!}{\includegraphics{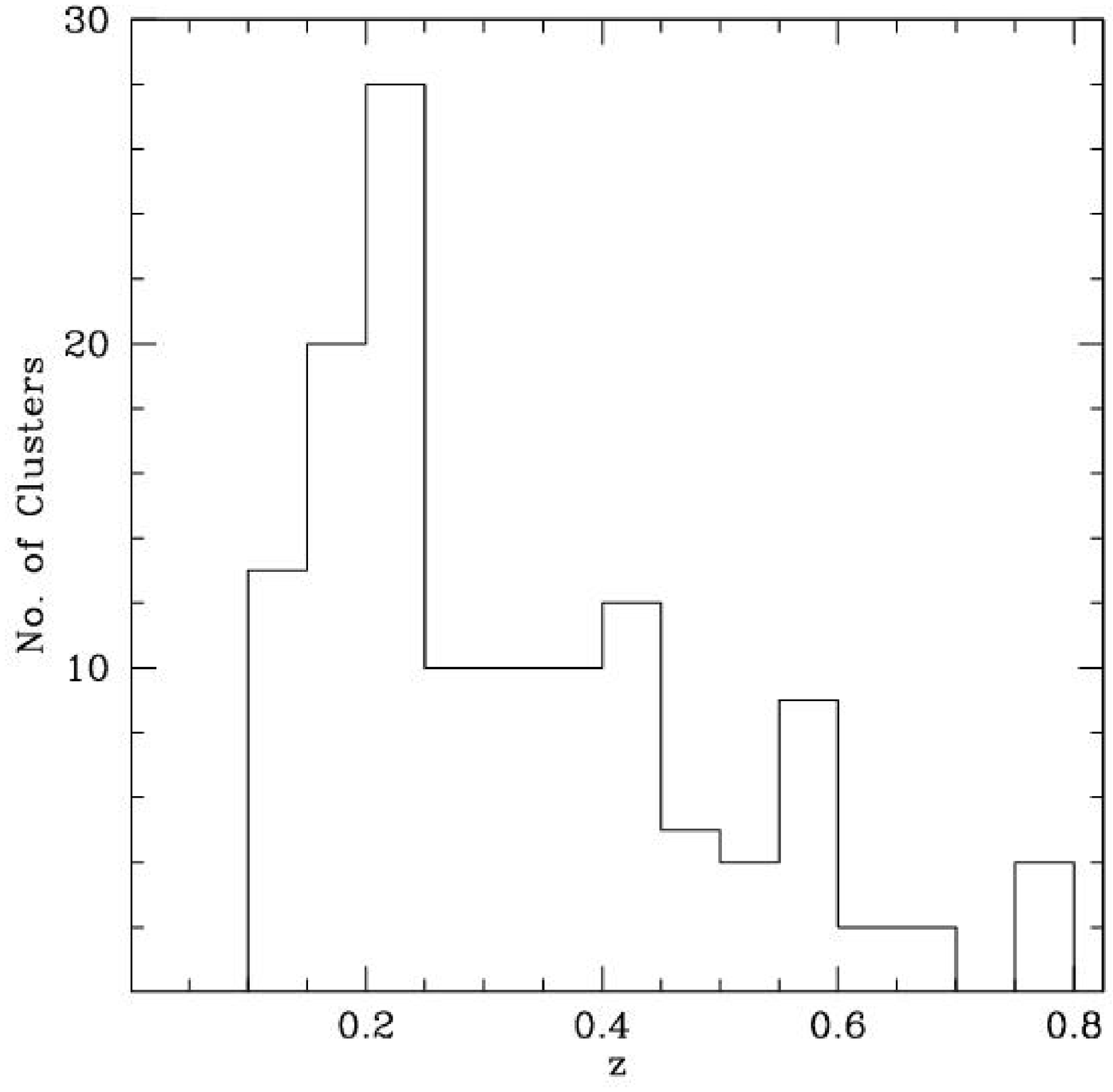}}
\end{center}
\figcaption{Histogram illustrating the number of clusters in the
sample as a function of redshift. \label{fig:cluszfig}}
\end{inlinefigure}

\clearpage

\clearpage

\clearpage

\begin{inlinefigure}
\begin{center}
\resizebox{\textwidth}{!}{\includegraphics{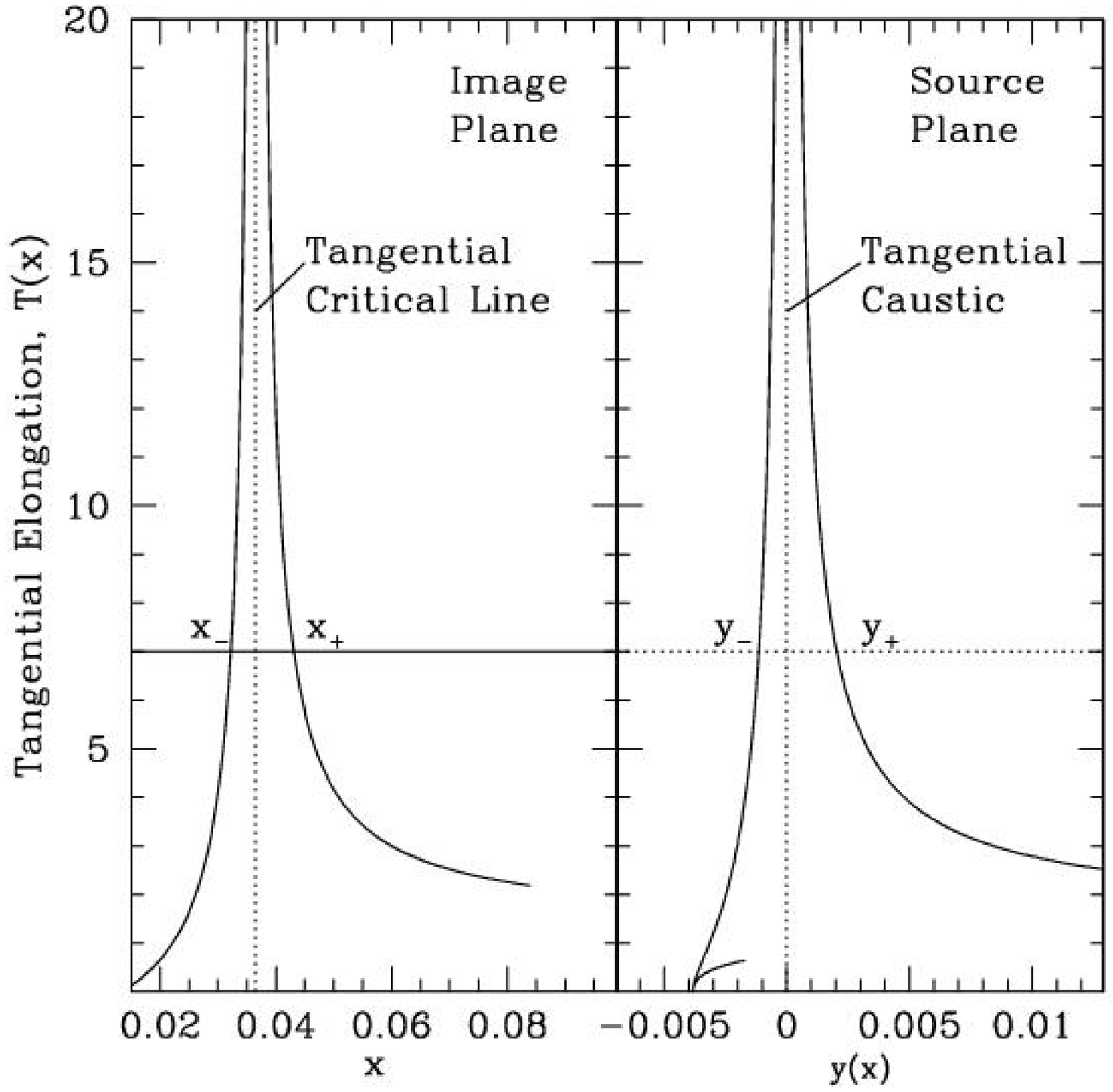}}
\end{center}
\figcaption{An illustration of how the tangential arc cross section is
found with a $1.0 \times 10^{15} M_{\odot}$ NFW ($\beta=1.0$) profile
at z=0.3 with a background at z=1.0.  The x and y positions
corresponding to a tangentially oriented axis ratio ($L/W$) greater
than 7. The value of $y_{-}$ or $y_{+}$ with the largest absolute
value is used in Eqn.~\ref{eq:sigtan} for calculating the
cross-section.
\label{fig:Tfig}}
\end{inlinefigure}

\clearpage

\begin{inlinefigure}
\begin{center}

\resizebox{\textwidth}{!}{\includegraphics{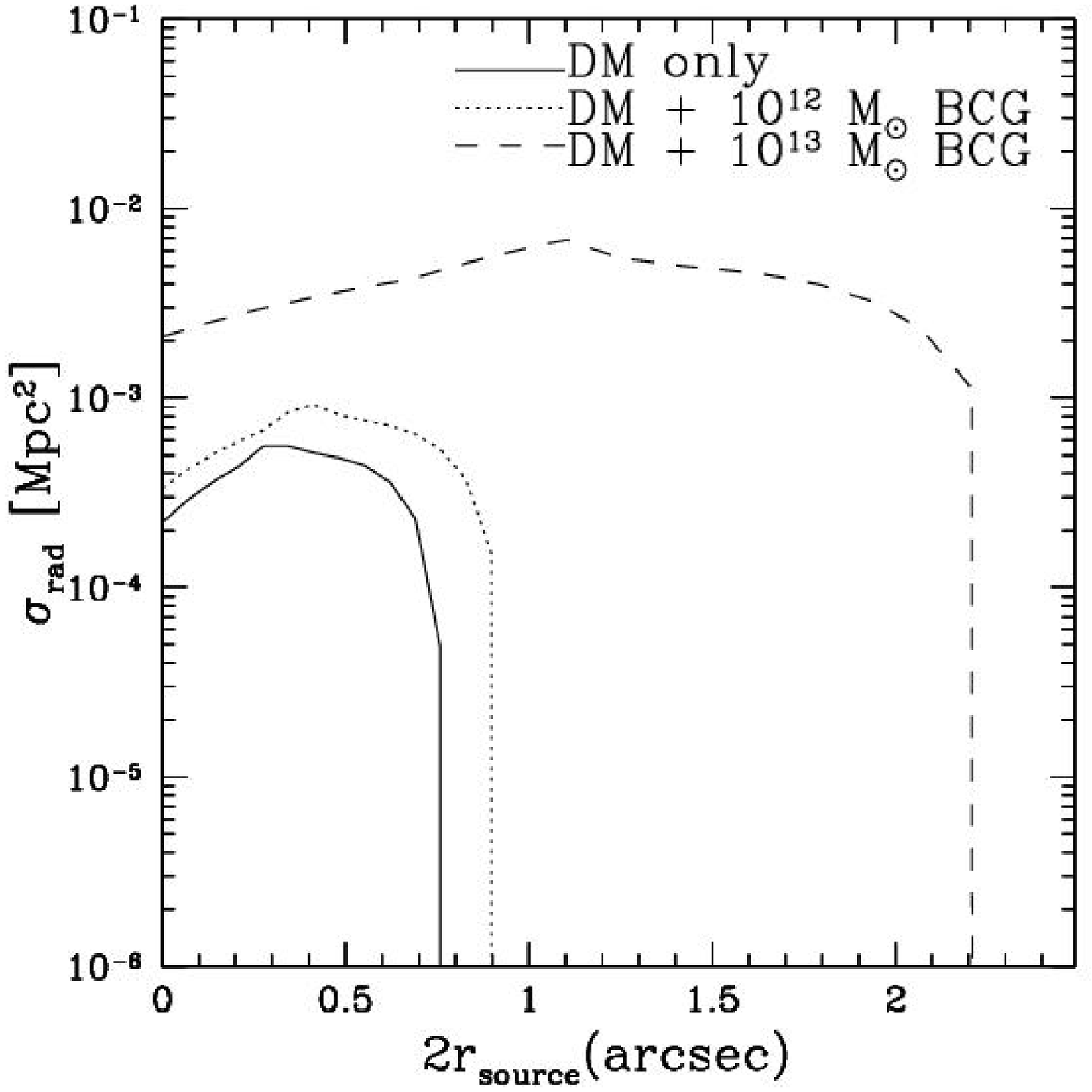}}
\end{center}

\figcaption{The radial arc cross section as a function of source
size. Shown are models with $1.0 \times 10^{15} M_{\odot}$ in DM
with profile slopes $\beta$=1.5 and various reasonable BCG masses
included. The lens redshift is at $z=0.2$ and the background
redshift is at $z=1.4$.  See \S~\ref{sec:massmod} for a
description of the mass models used.  Given the strong radial arc
cross section dependence, we adopt the z$\sim$1.4 size
distribution of Ferguson et al. (2004) taken from the GOODS
fields.  \label{fig:srcsz}}
\end{inlinefigure}

\clearpage

\begin{inlinefigure}
\begin{center}
\resizebox{\textwidth}{!}{\includegraphics{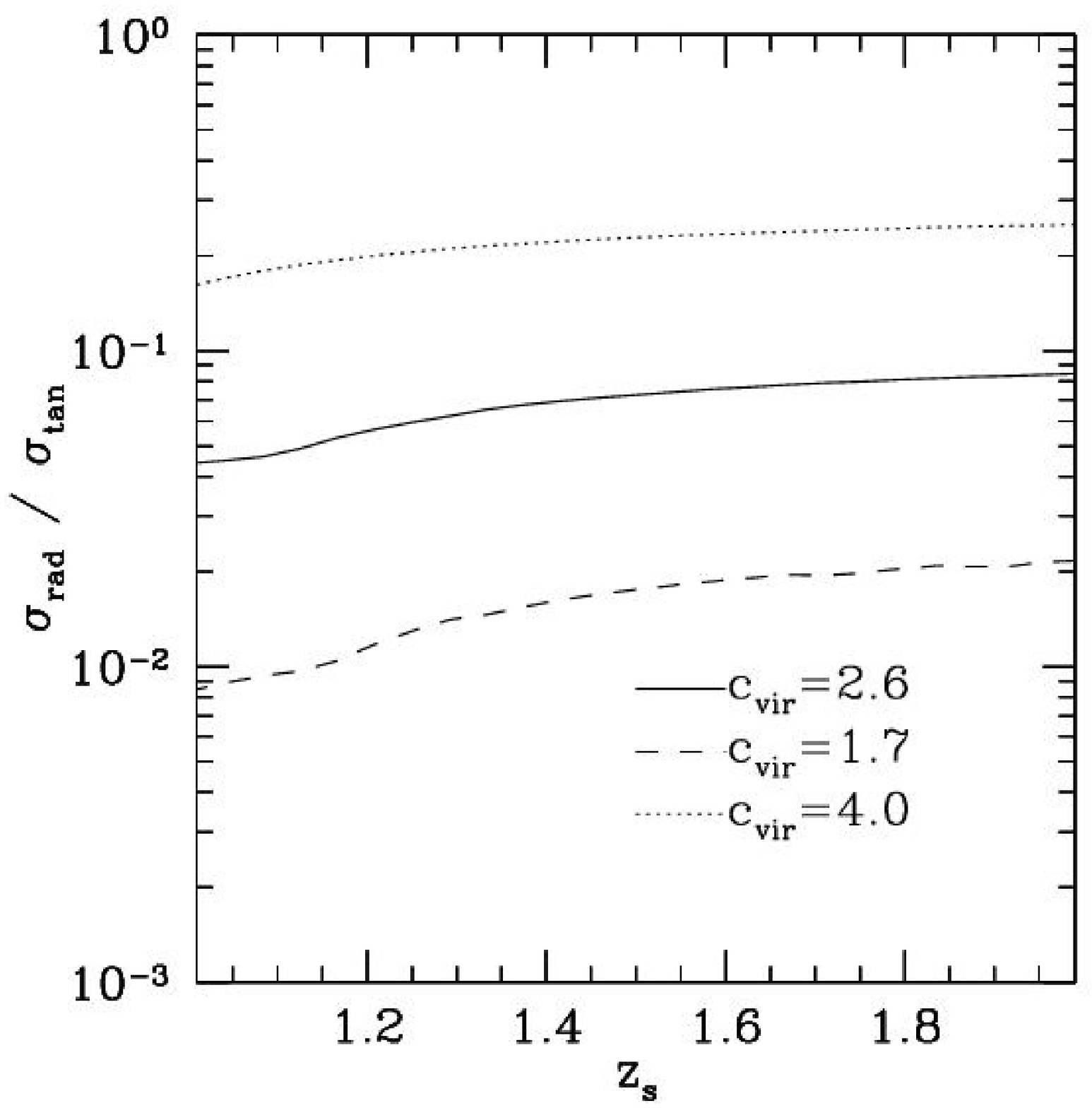}}
\end{center}

\figcaption{The radial to tangential cross section ratio as a
function of background source redshift while varying the
concentration parameter of the DM halo by $\pm$1-$\sigma$ as
prescribed by Bullock et al. 2001.  Shown are models with $1.0
\times 10^{15} M_{\odot}$ in DM with $\beta$=1.5 at a lens
redshift $z=0.2$.  The assumed radial arc source size distribution
is that found by Ferguson et al.  Note that the cross section
ratio is relatively constant as a function of background source
redshift.  \label{fig:ceffect}}
\end{inlinefigure}

\clearpage
\begin{inlinefigure}
\begin{center}
\resizebox{\textwidth}{!}{\includegraphics{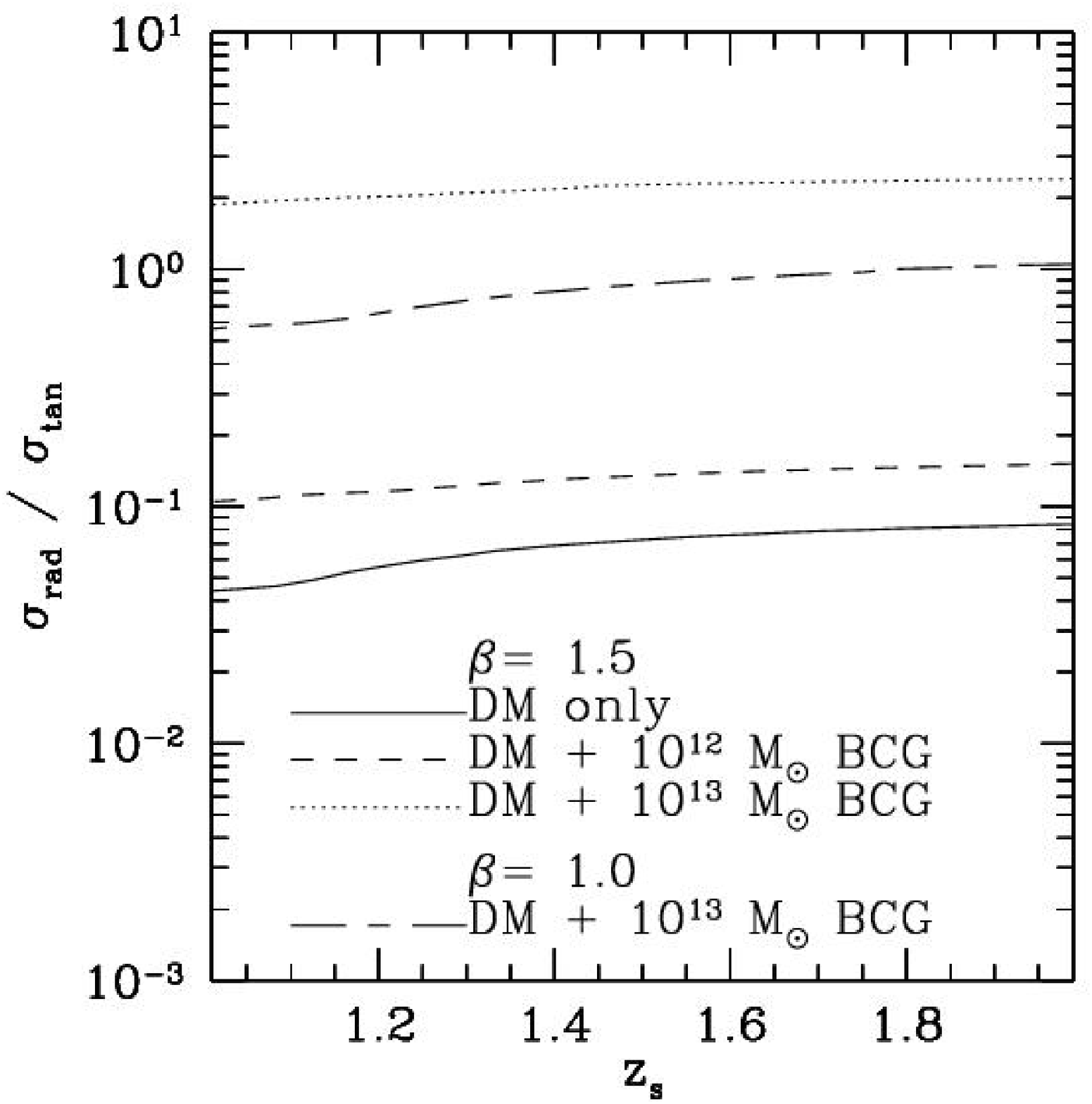}}
\end{center}
\figcaption{The radial to tangential arc cross section ratio as a
function of background source redshift while varying the BCG mass
and inner slope of the DM profile.  The assumed radial arc source
size distribution is that found by Ferguson et al. which we
assumed stays constant throughout the relevant source redshift
range.  Shown are models with $1.0 \times 10^{15} M_{\odot}$ in DM
at a lens redshift $z=0.2$.  Note that the $\beta$=1.0 and
$\beta$=1.5 models with a $10^{13} M_{\odot}$ BCG have expected
arc number ratios within a factor of few of each other.
\label{fig:bcgeffect}}
\end{inlinefigure}

\clearpage

\clearpage

\clearpage

\begin{table*}
\begin{center}
\caption{New Spectroscopic Observations\label{tab:spec}}
\begin{tabular}{lcccccc}
\tableline\tableline
Cluster &Date &Target &Instrument &Exposure &Redshift&Notes\\
&  & &  &time (ks) & &\\
\tableline

Abell 2259&July 27, 2001&Tan Arc& ESI&7.2&1.477&Arc A1\\
MS 1455&July 27, 2001&Rad Arc Cand&ESI&4.5&-&Not lensed\\
Abell 370&Oct 19, 2001&Rad Arc&LRIS&7.2&-&No detection\\
Abell 1835&April 11, 2002&Rad Arc&ESI&1.8&-&No detection\\
Abell 963&Nov 22, 2003&Tan Arc&LRIS&7.2&1.958&Arc H1\\
GC0848+44&Feb 22, 2004&Rad Arc& LRIS&3.0&-&Faint Cont.\\
Abell 773&Feb 22, 2004&Tan Arc& LRIS&5.4&1.114&Arc F11\\
GC 1444&July 19, 2004&Rad Arc&ESI&7.2&1.151&Arc A1\\
3c435a&July 19, 2004&Rad Arc Cand&ESI&5.4&-&Not lensed\\
Abell 2667&July 19, 2004&Rad Arc Cand&ESI&3.6&-&Not lensed\\
Abell 2667&July 19, 2004&Tan Arc&ESI&3.6&1.034&Arc A1\\
AC 118&July 19,20, 2004&2 Rad Arc&ESI&4.6/7.2&-&Faint Cont.\\
MS 0440&Dec 12, 2004&Rad Arc&LRIS&5.4&-&No detection; arc A17\\
IRAS 0910&Dec 12, 2004&Rad Arc&LRIS&0.6&-&Not lensed\\
3c220&Dec 12, 2004&Rad Arc&LRIS&2.4&-&Not lensed\\
R0451&Dec 13, 2004&Tan Arc&LRIS&3.6&2.007&Arc A1\\
RCS0224&Dec 13, 2004&Rad Arc&LRIS&5.4&1.050&Arc R1\\ 
\tableline
\end{tabular}
\end{center}
\end{table*}

\clearpage
\begin{deluxetable}{lccccc}

\tablecolumns{6}
\tablewidth{0pc}

\tablecaption{Summary of Giant Arcs with $L/W>7$\label{tab:arcsum}}
\tablehead{\colhead{Cluster}    &  \colhead{No. of}    & \colhead{Tangential} & \colhead{Radial} & \colhead{Ratio R/T}&\colhead{68\% Confidence Range}\\
\colhead{Sample} & \colhead{Clusters}}
\startdata
Edge & 44&15 & 4&0.27&0.13-0.53\\
Smith & 10& 38& 1 & 0.03&0.005-0.09\\
EMSS & 12&13 & 2 &0.15&0.05-0.40\\
Total & 128&104 & 12 &0.12&0.08-0.16\\
\enddata
\end{deluxetable}

\clearpage

\begin{deluxetable}{lccccccc}
\tabletypesize{\small}
\tablecolumns{10}
\tablecaption{WFPC2 Cluster Catalog\label{fig:clustable}}
\tablehead{
\colhead{Cluster} & \colhead{PID} & \colhead{$z_{clus}$}& \colhead{$\alpha$}& \colhead{$\delta$}&\colhead{Exp.}& \colhead{Filter} & \colhead{Cluster}\\
\colhead{}&\colhead{}&\colhead{}&\colhead{(J2000.0)}&\colhead{(J2000.0)}&\colhead{time (ks)}&\colhead{}&\colhead{Sample}}
\startdata
A11\tablenotemark{\dagger}&8719&0.166&00 12 33.9&-16 28 06.9&0.8&606&1\\
AC118\tablenotemark{\dagger}&5701&0.308&00 14 20.6&-30 24 01.5&6.5&702&..\\
MS0016&5378&0.541&00 18 33.6&+16 26 16.0&12.6/16.8&555/814&3\\
Cl0024\tablenotemark{\dagger}\tablenotemark{\ddagger}&5453&0.39&00 26 35.5&+17 09 50.7&23.4/19.8&450/814&..\\
A68\tablenotemark{\ddagger}&8249&0.255&00 37 06.8&+09 09 23.4&7.5&702&2\\
3c16&6675&0.405&00 37 45.4&+13 20 09.8&7.1&702&..\\
GC0054\tablenotemark{\ddagger}&5378&0.560&00 56 56.9&-27 40 29.6&12.6/16.8&555/814&..\\
PKS0101&6675&0.390&01 04 24.1&+02 39 43.4&7.5&702&..\\
A209\tablenotemark{\ddagger}&8249&0.206&01 31 52.6&-13 36 40.8&7.8&702&2\\
A267\tablenotemark{\ddagger}&8249&0.230&01 52 41.9&+01 00 25.9&7.5&702&2\\
A291\tablenotemark{\dagger}&8301&0.196&2 01 39.90&-02 11 39.7&0.8&606&1\\
GC0210&8131&0.270&02 10 26.0&-39 29 42.9&7.8&702&..\\
5c6.124\tablenotemark{\ddagger}&6675&0.448&02 16 40.9&+32 50 47.1&10.4&702&..\\
RCS0224\tablenotemark{\dagger}\tablenotemark{\ddagger}&9135&0.77&02 24 30.82&-00 02 27.8&13.2/6.6&606/814&..\\
GC0231&6000&0.607&02 31 42.7&+00 48 41.0&15.3&606&..\\
A370\tablenotemark{\dagger}\tablenotemark{\ddagger}&6003&0.375&02 39 53.1&-01 34 54.8&5.6&675&..\\
A383\tablenotemark{\dagger}\tablenotemark{\ddagger}&8249&0.189&02 48 03.3&-03 31 44.4&7.5&702&2\\
GC030518&5991&0.42&03 05 18.1&+17 28 24.9&2.1/2.4&606/814&3\\
GC03053\tablenotemark{\ddagger}&5991&0.43&3 05 31.6&+17 10 03.1&2.1/2.4&606/814&3\\
GC0303&5378&0.420&03 06 19.0&+17 18 49.6&12.6&702&..\\
Cl0317&7293&0.583&03 20 00.8&+15 31 50.1&2.6/2.5&555/814&..\\
GC0329&8131&0.45&03 29 02.8&+02 56 23.3&10.4&702&..\\
GC0337&7374&0.59&03 37 45.1&-25 22 35.8&11.0&702&..\\
GC0341&8131&0.44&03 41 59.1&-44 59 58.3&11.2&702&..\\
GC0412&5378&0.51&04 12 52.1&-65 50 48.5&12.6/14.7&555/814&..\\
RXJ0439\tablenotemark{\ddagger}&8719&0.245&04 39 00.5&+07 16 09.5&04.1&606&1\\
MS0440\tablenotemark{\dagger}\tablenotemark{\ddagger}&5402&0.190&04 43 09.7&+02 10 19.5&22.2&702&3\\
RXJ0451\tablenotemark{\ddagger}&8719&0.430&04 51 54.6&+00 06 19.3&1.0&606&1\\
MS0451\tablenotemark{\dagger}\tablenotemark{\ddagger}&5987&0.55&04 54 10.6&-03 00 50.7&10.4&702&3\\
GC0720&7374&0.268&07 20 17.8&+71 32 13.4&4.8&702&..\\
GC072056&8131&0.230&07 20 53.8&+71 08 59.3&5.2&702&..\\
A586&8301&0.170&07 32 20.3&+31 38 00.1&0.8&606&1\\
PKS0745\tablenotemark{\ddagger}&7337&0.103&07 47 31.3&-19 17 40.0&2.1/1.8&555/814&..\\
GC0818&8325&0.260&08 18 57.3&+56 54 24.5&7.2&702&..\\
GC0819&8325&0.226&08 19 18.3&+70 55 04.6&6.9&702&..\\
RXJ0821&8301&0.109&08 21 02.3&+07 51 47.3&0.6&606&1\\
A646&8301&0.127&08 22 09.6&+47 05 52.5&0.6&606&1\\
4c+55\tablenotemark{\ddagger}&8719&0.242&08 34 54.9&+55 34 21.3&1.0&606&1\\
GC0841+64&7374&0.342&08 41 07.6&+64 22 25.7&7.2&702&..\\
GC0841&8325&0.235&08 41 44.1&+70 46 53.2&6.9&702&..\\
A697\tablenotemark{\ddagger}&8301&0.282&08 42 57.6&+36 21 59.1&1.0&606&1\\
GC0848\tablenotemark{\dagger}&7374&0.570&08 48 48.0&+44 56 16.7&12.0&702&..\\
GC0848N\tablenotemark{\ddagger}&7374&0.543&08 48 49.3&+44 55 48.2&12.0&702&..\\
GC0849&8325&0.240&08 49 10.8&+37 31 08.1&7.8&702&..\\
Z2089&8301&0.235&09 00 36.8&+20 53 39.6&1.0&606&1\\
3c215&5988&0.411&09 06 31.8&+16 46 11.7&7.8/5.0&555/814&..\\
RXJ0911&8705&0.77&09 11 26.5&+05 50 14.5&12.5/12.5&606/814&..\\
IRAS0910&6443&0.442&09 13 45.5&+40 56 27.9&4.34&814&..\\
A773\tablenotemark{\ddagger}\tablenotemark{\dagger}&8249&0.217&09 17 53.5&+51 44 1.0&7.2&702&2\\
A795&8301&0.136&09 24 05.3&+14 10 21.0&0.6&606&1\\
3c220\tablenotemark{\ddagger}&6778&0.620&09 32 40.1&+79 06 28.9&11.4/11.5&555/814&..\\
GC0939&5378&0.407&09 43 03.0&+46 56 33.3&4.0/6.3&555/814&..\\
GC0943&6581&0.70&09 43 42.7&+48 05 03.1&20.2&702&..\\
A868\tablenotemark{\ddagger}&8203&0.153&09 45 26.4&-08 39 06.6&4.4&606&..\\
Z2701&8301&0.215&09 52 49.1&+51 53 05.2&1.0&606&1\\
GC0952&6478&0.377&09 52 56.0&+43 55 28.8&7.0&555/814&..\\
GC0957&5979&0.390&10 01 20.9&+55 53 50.9&32.2/2.3&555/814&..\\
A963\tablenotemark{\ddagger}&8249&0.206&10 17 03.7&+39 02 49.2&7.8&702&2\\
A980&8719&0.158&10 22 28.4&+50 06 19.9&0.8&606&1\\
Z3146\tablenotemark{\ddagger}&8301&0.291&10 23 39.6&+04 11 10.8&1.0&606&1\\
A990&8301&0.142&10 23 39.9&+49 08 37.9&0.6&606&1\\
Z3179\tablenotemark{\ddagger}&8301&0.143&10 25 58.0&+12 41 07.5&0.6&606&1\\
A1033&8301&0.126&10 31 44.3&+35 02 29.0&0.6&606&1\\
A1068&8301&0.139&10 40 44.4&+39 57 10.9&0.6&606&1\\
A1084\tablenotemark{\dagger}&8301&0.133&10 44 32.9&-07 04 08.0&0.6&606&1\\
5c210&5988&0.478&10 52 36.1&+48 40 01.2&10.8/5.3&555/814&..\\
A1201\tablenotemark{\ddagger}&8719&0.151&11 12 54.5&+13 26 08.0&0.8&606&1\\
A1204&8301&0.171&11 13 20.5&+17 35 41.0&0.8&606&1\\
Z3916&8301&0.204&11 14 21.8&+58 23 19.8&1.0&606&1\\
A1246&8301&0.190&11 23 58.8&+21 28 46.2&0.8&606&1\\
RXJ1133\tablenotemark{\dagger}\tablenotemark{\ddagger}&8719&0.394&11 33 13.2&+50 08 40.8&1.0&606&1\\
MS1137\tablenotemark{\ddagger}&6668/5987&0.782&11 40 22.3&+66 08 14.1&13.8/14.4&606/814&3\\
A1366&8719&0.116&11 44 36.9&+67 24 20.4&0.6&606&1\\
A1423&8719&0.213&11 57 17.3&+33 36 40.1&1.0&606&1\\
GC1205&8131&0.35&12 05 51.3&+44 29 08.9&7.8&702&..\\
Z5247&8719&0.229&12 34 17.6&09 45 58.6&1.0&606&1\\
GC1256&8131&0.232&12 56 02.3&+25 56 36.7&4.4&702&..\\
5c12.251&6675&0.312&13 05 51.7&+36 39 27.3&5.0&702&..\\
A1682\tablenotemark{\dagger}\tablenotemark{\ddagger}&8719&0.221&13 06 45.8&+46 33 30.4&1.0&606&1\\
3c281&5988&0.600&13 07 54.0&+06 42 14.5&9.6/7.6&606/814&..\\
GC1309&8325&0.290&13 09 56.2&+32 22 13.0&7.8&702&..\\
A1689\tablenotemark{\ddagger}\tablenotemark{\dagger}&6004&0.183&13 11 29.4&-01 20 28.7&44.2/6.0&555/814&..\\
GHO1322&6278&0.755&13 24 47.2&+30 59 00.1&15.8/15.8&606/814&..\\
GC1322&5234&0.571&13 24 48.9&30 11 39.3&8.0/16.0&606/814&..\\
GC1335&8131&0.382&13 34 57.6&+37 50 29.9&7.8&702&..\\
A1763\tablenotemark{\ddagger}&8249&0.223&13 35 20.2&+41 00 04.6&7.8&702&2\\
GC1347&8131&0.470&13 48 00.9&+07 52 23.7&10.4&702&..\\
MS1358\tablenotemark{\ddagger}&5989&0.33&13 59 50.5&+62 31 06.8&3.6&606/814&3\\
A1835\tablenotemark{\ddagger}\tablenotemark{\dagger}&8249&0.253&14 01 02.1&+02 52 42.3&7.5&702&2\\
GC1409&5378&0.460&14 11 20.5&+52 12 09.6&12.6&702&..\\
A1902&8719&0.160&14 21 40.4&+37 17 30.5&0.8&606&1\\
A1914\tablenotemark{\ddagger}&8301&0.170&14 25 56.7&+37 48 58.08&0.8&606&1\\
GC1444\tablenotemark{\dagger}&8325&0.298&14 44 06.8&+63 44 59.6&7.5&702&..\\
GC1446&5707&0.37&14 49 31.2&+26 08 36.8&4.4&702&..\\
A1978&8719&0.147&14 51 09.4&+14 36 43.7&0.8&606&1\\
MS1455\tablenotemark{\ddagger}&8301&0.258&14 57 15.1&+22 20 34.9&1.0&606&1,3\\
A2009&8301&0.153&15 00 19.5&+21 22 10.7&0.8&606&1\\
MS1512\tablenotemark{\ddagger}&6832/6003&0.372&15 14 22.4&+36 36 21.0&10.4/5.8/19.8&555/675/814&3\\
A2125&7279&0.247&15 41 02.0&+66 16 26.2&2.6/2.6&606/814&..\\
A2146&8301&0.234&15 56 13.9&+66 20 52.5&1.0&606&1\\
GC1601&5378&0.539&16 03 13.0&+42 45 50.7&16.8&702&..\\
RXJ1621.4&8719&0.465&16 21 24.8&+38 10 09.0&1.0&606&1\\
MS1621\tablenotemark{\dagger}\tablenotemark{\ddagger}&6825&0.426&16 23 35.2&+26 34 28.2&4.6/4.6&555/814&3\\
A2204\tablenotemark{\dagger}&8301&0.151&16 32 46.9&+05 34 33.1&0.8&606&1\\
GC1633&7374&0.239&16 33 42.1&+57 14 13.9&4.8&702&..\\
A2218\tablenotemark{\ddagger}&5701/7343&0.176&16 35 49.3&+66 12 43.5&8.4/6.5&606/702&2\\
A2219\tablenotemark{\ddagger}\tablenotemark{\dagger}&6488&0.225&16 40 19.8&+46 42 41.9&14.4&702&2\\
GC1648&8131&0.377&16 48 42.5&+60 19 09.7&7.8&702&..\\
GC1701&8325&0.220&17 01 47.7&+64 21 00.5&7.5&702&..\\
GC1702&8325&0.224&17 02 13.9&+64 19 54.2&7.5&702&..\\
A2254&8301&0.178&17 17 45.9&+19 40 49.1&0.8&606&1\\
Z8197&8301&0.114&17 18 12.1&+56 39 56.0&0.6&606&1\\
A2259\tablenotemark{\ddagger}&8719&0.164&17 20 9.7&+27 40 07.4&0.8&606&1\\
A2261\tablenotemark{\ddagger}&8301&0.224&17 22 27.2&+32 07 57.5&1.0&606&1\\
A2294\tablenotemark{\dagger}\tablenotemark{\ddagger}&8301&0.178&17 24 12.6&+85 53 11.6&0.8&606&1\\
RXJ1750&8301&0.171&17 50 16.9&+35 04 58.7&0.8&606&1\\
MS2053\tablenotemark{\ddagger}&5991/6745&0.58&20 56 21.4&-04 37 50.9&3.3/2.4/3.2&606/702/814&3\\
AC103&5701&0.311&20 57 01.1&-64 39 47.2&6.5&702&..\\
3c435a&6675&0.471&21 29 05.5&+07 33 00.2&12.8&702&..\\
RXJ2129&8301&0.235&21 29 40.0&+00 05 20.7&1.0&606&1\\
MS2137\tablenotemark{\ddagger}\tablenotemark{\dagger}&5991/5402&0.313&21 40 14.9&-23 39 39.5&2.4/2.6/22.2&606/814/702&3\\
A2390\tablenotemark{\ddagger}&5352&0.228&21 53 36.9&+17 41 43.4&8.4/10.5&555/814&..\\
GC2157&6581&0.70&21 57 50.6&+03 48 47.3&18.8&702&..\\
A2409&8719&0.147&22 00 52.6&+20 58 09.7&0.8&606&1\\
Cl2244\tablenotemark{\ddagger}&5352&0.330&22 47 12.2&-02 05 38.6&8.4/12.6&555/814&..\\
AC114\tablenotemark{\dagger}\tablenotemark{\ddagger}&7201/5935&0.312&22 58 48.3&-34 48 07.2&15.6/16.6&702/702&..\\
4c27.51&6675&0.319&23 25 00.4&+28 03 11.4&5.2&702&..\\
A2667\tablenotemark{\ddagger}&8882&0.2264&23 51 39.4&-26 05 03.8&12.0/4.0/4.0&450/606/814&..
\enddata
\tablenotetext{\dagger}{Cluster has a radial arc candidate}
\tablenotetext{\ddagger}{Cluster has a tangential arc candidate with $L/W>7$}
\tablenotetext{1}{Cluster belongs to Edge Sample}
\tablenotetext{2}{Cluster belongs to Smith Sample}
\tablenotetext{3}{Cluster belongs to EMSS Cluster Sample}
\end{deluxetable}

\clearpage

\begin{deluxetable}{lccccccccc}
\tabletypesize{\scriptsize}
\tablecolumns{10}
\tablecaption{Arc List\label{fig:arctable}}
\tablehead{
\colhead{Cluster} & \colhead{Arc Label} & \colhead{$z_{arc}$}& \colhead{z ref.}& \colhead{Mag}&\colhead{Filter}& \colhead{L/W} & \colhead{$\Delta$ N"}& \colhead{$\Delta$ E"}&\colhead{Type}}
\startdata
A11&A1&..&..&22.24$\pm$0.11&606&3.5$\pm$0.2&-0.7&-1.7&R\\
AC118&A1&..&..&24.29$\pm$0.19&702&9.4$\pm$1.2&2.0&-3.9&R\\
 &A2&..&..&24.55$\pm$0.27&702&12.1$\pm$1.4&-5.4&0.5&R\\
 &A3&..&..&24.33$\pm$0.21&702&10.4$\pm$0.9&-17.5&22.5&R\\
CL0024&A\tablenotemark{\dagger}&1.675&1&21.30$\pm$0.02,20.32$\pm$0.02&450/814&4.4$\pm$0.2&-4.1&32.9&T\\
 &B\tablenotemark{\dagger}&1.675&1&22.19$\pm$0.02,21.61$\pm$0.03&450/814&3.7$\pm$0.7&-17.7&29.4&T\\
 &C&1.675&1&21.08$\pm$0.01,19.94$\pm$0.01&450/814&5.2$\pm$0.2&-29.0&19.8&T\\
 &D&1.675&1&22.13$\pm$0.03,21.35$\pm$0.03&450/814&4.6$\pm$0.1&11.9&-17.8&T\\
 &E\tablenotemark{\dagger}&1.675&1&23.24$\pm$0.03,20.98$\pm$0.01&450/814&2.6$\pm$0.1&-0.7&-3.9&R\\
 &A1&..&..&22.25$\pm$0.03,22.18$\pm$0.05&450/814&11.39$\pm$2.4&22.5&-30.0&T\\
 &A2&..&..&23.47$\pm$0.06,23.24$\pm$0.11&450/814&12.2$\pm$1.2&-47.3&-21.3&T\\
 &A3&..&..&23.63$\pm$0.06,22.17$\pm$0.04&450/814&8.6$\pm$0.5&-52.1&6.8&T\\
 &A4\tablenotemark{\dagger}&..&..&24.43$\pm$0.07,22.19$\pm$0.02&450/814&2.3$\pm$0.1&-0.9&6.4&R\\
 &A5&..&..&24.28$\pm$0.08,22.38$\pm$0.03&450/814&7.5$\pm$0.9&-2.7&60.9&T\\
A68&C0ab\tablenotemark{\dagger}&1.60&2&20.94$\pm$0.04&702&9.6$\pm$0.7&2.3&7.7&T\\
 &C0c&1.60&2&23.99$\pm$0.19&702&8.3$\pm$0.4&-15.9&-11.7&T\\
 &C4&2.625&3&22.82$\pm$0.07&702&6.2$\pm$0.6&-18.6&12.1&T\\
 &C6/C20&..&..&21.77$\pm$0.05&702&14.1$\pm$1.4&27.4&-32.3&T\\
 &C8&0.861&4&22.73$\pm$0.04&702&4.0$\pm$0.2&30.3&-46.6&T\\
 &C9&..&..&23.78$\pm$0.14&702&7.2$\pm$0.4&37.4&-38.7&T\\
 &C12\tablenotemark{\dagger}&1.265&4&21.09$\pm$0.02&702&4.8$\pm$0.1&56.6&-28.8&T\\
 &C18&..&..&22.75$\pm$0.05&702&13.6$\pm$2.5&5.5&24.2&T\\
GC0054&A1&..&..&24.18$\pm$0.07;23.47$\pm$0.07&555/814&7.4$\pm$0.1&-3.2&14.1&T\\
A209&D1&..&..&21.60$\pm$0.02&702&8.0$\pm$0.3&15.1&-16.4&T\\
A267&E0&..&..&23.90$\pm$0.17&702&8.7$\pm$0.9&23.3&12.6&T\\
A291&A1&..&..&23.14$\pm$0.15&606&3.8$\pm$0.4&-1.2&0.2&R\\
5c6.124&A1\tablenotemark{\dagger}&..&..&23.27$\pm$0.05&702&8.3$\pm$0.4&1.40&1.5&T\\
RCS0224&C&4.879&12&24.22$\pm$0.24,22.11$\pm$0.06&606/814&15.6$\pm$3.7&-8.9&-12.4&T\\
 &R1&1.055&7&24.25$\pm$0.10,23.09$\pm$0.06&606/814&5.3$\pm$0.9&-9.5&5.5&R\\
 &A1\tablenotemark{\dagger}&..&..&24.26$\pm$0.10,24.94$\pm$0.4&606/814&5.3$\pm$1.8&-1.5&-0.8&R\\
 &A2\tablenotemark{\dagger}&..&..&22.94$\pm$0.05,22.15$\pm$0.04&606/814&9.6$\pm$0.4&-6.1&7.9&T\\
 &A3&..&..&24.50$\pm$0.2,22.56$\pm$0.06&606/814&8.6$\pm$1.2&-14.5&-6.3&T\\
A370&A0\tablenotemark{\dagger}&0.724&25&18.92$\pm$0.01&675&13.6$\pm$0.6&-47.3&3.7&T\\
 &A1&..&..&22.91$\pm$0.13&675&32.4$\pm$12.2&18.0&-14.7&T\\
 &A2&..&..&23.13$\pm$0.13&675&19.6$\pm$4.9&21.8&-4.9&T\\
 &B2&0.806&26&22.93$\pm$0.06&675&4.7$\pm$0.1&-17.0&-3.9&T\\
 &B3&0.806&26&22.99$\pm$0.06&675&4.4$\pm$0.3&-17.9&-9.1&T\\
 &R1/R2\tablenotemark{\dagger}&..&..&22.22$\pm$0.05&675&7.0$\pm$0.4&-31.2&0.4&R\\
A383&B0a/B1abc/B4abc&1.01&5,6&20.22$\pm$0.02&702&16.3$\pm$0.9&-10.4&-12.0&T\\
 &B0b/B1d\tablenotemark{\dagger}&1.01&5&22.69$\pm$0.11&702&20.2$\pm$3.3&2.6&1.8&R\\
 &B2ab/B2c/B3a\tablenotemark{\dagger}&..&..&22.10$\pm$0.05&702&17.4$\pm$1.5&-22.5&-4.3&T\\
 &B5&..&..&23.43$\pm$0.24&702&16.4$\pm$1.9&-15.6&8.7&T\\
 &B6&..&..&23.10$\pm$0.10&702&9.1$\pm$1.2&-6.4&13.9&T\\
 &B7&..&..&24.36$\pm$0.18&702&8.6$\pm$0.5&4.3&-18.7&T\\
 &B11&..&..&24.33$\pm$0.20&702&7.5$\pm$0.4&7.1&22.9&T\\
GC03053&A1&..&..&21.93$\pm$0.09,21.03$\pm$0.08&606/814&20.9$\pm$1.2&17.03&-6.39&T\\
GC0337&A1\tablenotemark{\dagger}&..&..&23.65$\pm$0.08&702&9.3$\pm$0.7&-7.1&6.7&T\\
MS0440&A1\tablenotemark{\dagger}&0.5317&7&21.31$\pm$0.01&702&2.2$\pm$0.1&-7.3&18.9&T\\
 &A3\tablenotemark{\dagger}&..&..&22.51$\pm$0.01&702&10.2$\pm$0.4&10.1&-20.4&T\\
 &A16\tablenotemark{\dagger}&..&..&20.67$\pm$0.01&702&2.9$\pm$0.1&4.8&4.7&R\\
 &A17\tablenotemark{\dagger}&..&..&20.26$\pm$0.01&702&3.2$\pm$0.1&5.9&0.7&R\\
RXJ0451&A1&2.007&7&20.24$\pm$0.03&606&19.7$\pm$3.7&-3.8&38.0&T\\
 &A2&..&..&22.82$\pm$0.13&606&10.3$\pm$1.7&21.6&-18.6&T\\
MS0451&A1&2.91&13&22.33$\pm$0.04&702&7.6$\pm$0.3&-4.2&31.5&T\\
 &A3\tablenotemark{\dagger}&..&..&24.48$\pm$0.20&702&9.2$\pm$1.4&-6.5&11.3&T\\
 &A4&..&..&22.48$\pm$0.03&702&7.2$\pm$0.2&11.9&-18.0&T\\
 &A5\tablenotemark{\dagger}&..&..&25.21$\pm$0.32&702&11.4$\pm$2.8&-1.7&1.7&R\\
 &A6&..&..&23.66$\pm$0.08&702&4.6$\pm$0.4&44.2&81.9&R\\
PKS0745&A&0.433&14&20.73$\pm$0.02;18.96$\pm$0.01&555/814&4.3$\pm$0.1&5.8&-18.3&T\\
4c+55&A1&..&..&21.04$\pm$0.07&606&8.9$\pm$2.1&-13.0&-7.1&T\\
A697&A1&..&..&22.79$\pm$0.23&606&8.3$\pm$1.9&-20.6&-9.0&T\\
GC0848N&A/B/C&3.356&15&22.71$\pm$0.04&702&8.8$\pm$0.7&2.0&-5.8&T\\
GC0848&A1&..&..&25.20$\pm$0.20&702&4.7$\pm$0.6&-0.3&-0.9&R\\
A773&F0\tablenotemark{\dagger}&0.650&4&22.14$\pm$0.05&702&15.5$\pm$0.9&-19.8&-16.3&T\\
 &F3\tablenotemark{\dagger}&0.398&4&21.21$\pm$0.02&702&9.0$\pm$0.2&-17.7&-44.6&T\\
 &F4\tablenotemark{\dagger}&..&..&23.82$\pm$0.13&702&10.6$\pm$0.8&-30.2&-49.6&T\\
 &F9&..&..&21.58$\pm$0.03&702&9.8$\pm$0.3&-17.7&49.9&T\\
 &F11&1.114&7&22.86$\pm$0.06&702&9.5$\pm$0.5&13.0&48.1&T\\
 &F13&0.398&4&21.52$\pm$0.03&702&6.6$\pm$0.4&-8.1&59.5&T\\
 &F16&..&..&23.76$\pm$0.09&702&4.6$\pm$0.3&-4.6&-8.4&R\\
 &F18&0.487&4&23.39$\pm$0.11&702&10.0$\pm$0.8&-10.4&53.2&T\\
3c220&A1&1.49&16&22.30$\pm$0.04;21.36$\pm$0.03&555/814&10.1$\pm$0.1&8.6&0.4&T\\
A868&A1&..&..&22.28$\pm$0.05&606&16.8$\pm$2.1&10.3&17.1&T\\
 &A2&..&..&24.58$\pm$0.23&606&8.5$\pm$0.8&9.9&26.6&T\\
A963&H0&0.771&8&22.25$\pm$0.06&702&15.3$\pm$1.7&12.2&-0.3&T\\
 &H1/H2/H3&1.958\tablenotemark{\star}&7&21.73$\pm$0.06&702&30.2$\pm$4.0&-17.9&-1.5&T\\
 &H5&..&..&23.26$\pm$0.09&702&11.1$\pm$1.1&21.7&-4.8&T\\
 &A1&..&..&25.03$\pm$0.30&702&8.0$\pm$1.1&15.1&14.7&T\\
Z3146&A1&..&..&23.55$\pm$0.30&606&11.0$\pm$3.1&-10.6&25.4&T\\
Z3179&A1&..&..&24.72$\pm$1.20&606&$>20.0$&7.0&2.0&T\\
A1084&A1&..&..&23.81$\pm$0.58&606&20.3$\pm$6.2&0.5&-7.6&R\\
A1201&A1&0.451&9&20.35$\pm$0.03&606&7.8$\pm$0.7&1.8&-2.2&T\\
RXJ1133&A1&1.544&5&20.54$\pm$0.03&606&12.4$\pm$2.0&-0.7&10.7&T\\
 &A2&1.544&5&21.56$\pm$0.05&606&7.0$\pm$0.7&3.3&-1.2&R\\
MS1137&A1&..&..&24.28$\pm$0.10,23.58$\pm$0.10&606/814&8.2$\pm$0.9&1.6&15.6&T\\
 &A2&..&..&24.48$\pm$0.14,24.41$\pm$0.28&606/814&$>12.4$&17.9&4.8&T\\
 &A3&..&..&24.37$\pm$0.12,26.83$\pm$1.79&606/814&$>7.7$&-13.1&6.5&T\\
A1682&A1&..&..&25.17$\pm$1.15&606&8.3$\pm$3.5&3.3&-1.4&R\\
 &A2&..&..&22.28$\pm$0.11&606&25.3$\pm$5.7&3.8&-47.9&T\\
A1689&A1\tablenotemark{\dagger}&..&..&23.40$\pm$0.02,21.98$\pm$0.04&555/814&3.2$\pm$0.1&5.9&-14.3&R\\
 &16.2\tablenotemark{\dagger}&..&..&23.52$\pm$0.03,22.12$\pm$0.05&555/814&3.6$\pm$0.1&-0.5&-9.8&R\\
 &25.1\tablenotemark{\dagger}&..&..&24.18$\pm$0.05,22.51$\pm$0.07&555/814&4.4$\pm$0.2&-7.0&-14.8&R\\
 &5.1/5.2\tablenotemark{\dagger}&..&..&22.49$\pm$0.02,21.08$\pm$0.03&555/814&5.5$\pm$0.1&-18.7&-5.2&R\\
 &A2\tablenotemark{\dagger}&..&..&23.81$\pm$0.03,22.47$\pm$0.06&555/814&3.1$\pm$0.1&-21.0&0.8&R\\
 &A6\tablenotemark{\dagger}&..&..&23.04$\pm$0.03,22.47$\pm$0.06&555/814&6.9$\pm$0.1&9.6&9.8&R\\
 &19.1\tablenotemark{\dagger}&..&..&23.04$\pm$0.01,22.27$\pm$0.05&555/814&7.5$\pm$1.5&4.3&32.4&T\\
 &12.1&1.83&..&24.10$\pm$0.06,24.20$\pm$0.40&555/814&8.6$\pm$2.8&37.0&11.8&T\\
 &A3&..&..&24.10$\pm$0.05,24.02$\pm$0.32&555/814&10.0$\pm$3.8&45.9&-11.3&T\\
 &A4&..&..&23.09$\pm$0.02,22.84$\pm$0.13&555/814&15.3$\pm$3.5&44.8&-12.3&T\\
 &13.1/13.2/13.3&..&..&23.53$\pm$0.08,22.55$\pm$0.21&555/814&30.6$\pm$3.5&61.0&53.5&T\\
 &A5&..&..&24.25$\pm$0.09,22.72$\pm$0.13&555/814&7.6$\pm$1.9&-17.4&33.3&T\\
 &8.1/8.2/19.3/19.4&..&..&21.67$\pm$0.02,20.46$\pm$0.05&555/814&30.1$\pm$1.9&-31.5&36.7&T\\
 &1.1/2.1&3.05&..&22.35$\pm$0.02,21.51$\pm$0.06&555/814&13.8$\pm$1.9&31.7&-45.1&T\\
 &21.2\tablenotemark{\dagger}\tablenotemark{\ddagger}&..&..&-,23.85$\pm$0.34&555/814&$>12.5$&-16.6&18.9&R\\
 &29.2\tablenotemark{\ddagger}&..&..&25.97$\pm$0.39,-&555/814&13.9$\pm$1.0&53.5&8.1&T\\
 &A6\tablenotemark{a}&..&..&24.64$\pm$0.08,-&555/814&12.1$\pm$0.7&41.0&31.9&T\\
A1763&J1&..&..&24.87$\pm$0.28&702&8.7$\pm$1.5&-5.9&14.1&T\\
MS1358&B/C\tablenotemark{\ddagger}&4.92&17&22.19$\pm$0.11&606/814&14.9$\pm$1.6&-18.6&-11.7&T\\
A1835&K0\tablenotemark{\dagger}&..&..&24.10$\pm$0.13&702&6.1$\pm$0.3&7.2&-1.1&R\\
 &K2&..&..&23.45$\pm$0.11&702&7.7$\pm$0.7&-0.8&20.9&T\\
 &K3\tablenotemark{\dagger}&..&..&21.71$\pm$0.07&702&12.0$\pm$1.2&-20.8&23.3&T\\
A1914&A1&..&..&23.35$\pm$0.20&606&$>14.3$&-25.4&-11.8&T\\
GC1444&A1&1.151&7&23.88$\pm$0.11&702&4.4$\pm$0.5&1.2&0.8&R\\
MS1455&A1&..&..&21.66$\pm$0.08&606&15.6$\pm$6.6&8.9&17.9&T\\
MS1512&cB58\tablenotemark{\dagger}&2.72&18&20.62$\pm$0.01;20.40$\pm$0.02;19.93$\pm$0.01&555/675/814&4.1$\pm$0.2&4.6&-2.3&T\\
MS1621&A1\tablenotemark{\dagger}&..&..&25.19$\pm$0.52;25.42$\pm$0.23&555/814&$>4.1$&-1.5&-0.1&R\\
 &A2\tablenotemark{\dagger}&..&..&21.21$\pm$0.03;22.10$\pm$0.04&555/814&8.9$\pm$0.1&3.4&3.5&T\\
A2204&A1&..&..&22.36$\pm$0.10&606&5.9$\pm$0.4&7.9&2.7&R\\
 &A2&..&..&22.56$\pm$0.09&606&6.0$\pm$0.4&3.4&-0.3&R\\
A2218&M0bcd&0.702&23&21.01$\pm$0.01,20.16$\pm$0.01&606/702&11.0$\pm$0.7&-16.6&12.7&T\\
 &M4\tablenotemark{\ast}&1.034&23&20.25$\pm$0.02&606/702&15.3$\pm$1.0&-53.2&35.9&T\\
 &M1ab&2.515&24&21.33$\pm$0.02,20.82$\pm$0.02&606/702&6.5$\pm$0.7&22.4&0.9&T\\
 &M3ab&..&..&23.03$\pm$0.08,22.94$\pm$0.10&606/702&18.3$\pm$0.3&17.0&-17.6&T\\
 &730\tablenotemark{\ast}&..&..&22.02$\pm$0.05&702&17.2$\pm$0.2&-59.2&45.6&T\\
 &323&..&..&21.38$\pm$0.02,20.56$\pm$0.01&606/702&7.7$\pm$1.6&-14.3&24.2&T\\
 &382&..&..&23.75$\pm$0.16,25.47$\pm$1.00&606/702&16.2$\pm$2.0&35.1&2.7&T\\
 &A1&..&..&23.43$\pm$0.07,23.27$\pm$0.08&606/702&7.2$\pm$0.9&-38.6&9.1&T\\
 &H2/H3\tablenotemark{\ast}&..&..&23.19$\pm$0.12&702&22.9$\pm$3.0&-32.3&31.3&T\\
 &H1/273&..&..&21.73$\pm$0.01,21.89$\pm$0.03&606/702&8.9$\pm$1.5&-20.0&40.2&T\\
A2219&P0&1.070&3&21.58$\pm$0.02&702&10.0$\pm$0.6&-13.9&10.6&T\\
 &P2ab&2.730&3&22.38$\pm$0.05&702&26.9$\pm$3.0&17.3&-16.9&T\\
 &P2c&2.730&3&22.60$\pm$0.03&702&6.0$\pm$0.2&-3.8&-26.7&T\\
 &P13&..&..&23.94$\pm$0.11&702&9.0$\pm$0.7&4.6&-30.8&T\\
 &A1\tablenotemark{\dagger}&..&..&23.89$\pm$0.08&702&5.3$\pm$0.5&-19.2&31.1&R\\
A2259&A1&1.477&7&21.70$\pm$0.08&606&17.8$\pm$4.8&-0.7&10.7&T\\
A2261&A1\tablenotemark{\dagger}&..&..&22.82$\pm$0.20&606&25.5$\pm$8.1&-6.6&-26.3&T\\
 &A2\tablenotemark{\dagger}&..&..&21.76$\pm$0.04&606&7.7$\pm$0.9&8.4&23.5&T\\
A2294&A1&..&..&24.69$\pm$0.81&606&$>13.1$&2.3&6.7&R\\
 &A2&..&..&24.09$\pm$0.70&606&$>29.0$&-19.9&-24.9&T\\
MS2053&AB&3.146&29&21.67$\pm$0.03;21.27$\pm$0.05;20.90$\pm$0.05&606/702/814&11.6$\pm$0.7&14.6&-3.8&T\\
MS2137&A01/A02&1.501&10&21.76$\pm$0.06;21.91$\pm$0.04;21.54$\pm$0.12&606/702/814&13.6$\pm$5.7&15.4&1.5&T\\
 &AR&1.502&10&23.68$\pm$0.16;23.58$\pm$0.07;23.20$\pm$0.24&606/702/814&9.2$\pm$2.7&5.4&-0.2&R\\
A2390&A/C&1.033/0.913&22&21.51$\pm$0.03,19.9$\pm$0.02&555/814&10.7$\pm$0.5&17.7&-33.7&T\\
 &H3a&4.04&22&24.15$\pm$0.14,22.44$\pm$0.06&555/814&7.9$\pm$0.4&19.3&-45.1&T\\
 &H3b&4.04&22&24.46$\pm$0.12,22.88$\pm$0.06&555/814&4.1$\pm$0.1&9.4&-49.7&T\\
 &H5a&4.05&27&24.28$\pm$0.11,23.53$\pm$0.11&555/814&5.0$\pm$0.7&3.6&-20.4&T\\
 &H5b&4.05&27&24.21$\pm$0.11,22.82$\pm$0.06&555/814&5.9$\pm$0.9&-9.4&-24.6&T\\
 &A1&..&..&24.78$\pm$0.22,23.27$\pm$0.11&555/814&10.4$\pm$0.9&-6.3&5.8&T\\
 &A2\tablenotemark{\dagger}&..&..&-,20.76$\pm$0.04&555/814&19.1$\pm$2.1&-13.5&6.1&T\\
 &A3&..&..&22.59$\pm$0.10,22.99$\pm$0.11&555/814&8.1$\pm$0.5&-33.2&9.2&T\\
CL2244&A1\tablenotemark{\dagger}&2.237&11&20.62$\pm$0.02,20.00$\pm$0.02&555/814&10.8$\pm$0.1&1.3&-8.3&T\\
 &A4&..&..&23.74$\pm$0.13,24.72$\pm$0.59&555/814&10.4$\pm$1.9&-1.9&33.2&T\\
AC114&A4/A5&..&..&22.59$\pm$0.04&702&6.9$\pm$0.4&-2.9&-0.3&R\\
 &S1\tablenotemark{\dagger}&1.867&21&21.97$\pm$0.02&702&2.5$\pm$0.1&12.7&2.0&T\\
 &S2\tablenotemark{\dagger}&1.867&21&21.91$\pm$0.02&702&3.1$\pm$0.1&12.7&2.0&T\\
 &C1&2.854&19&24.17$\pm$0.10&702&5.6$\pm$0.2&29.0&5.6&T\\
 &T1&..&..&24.68$\pm$0.36&702&29.0$\pm$4.7&-24.7&2.5&T\\
 &T2&..&..&24.08$\pm$0.13&702&14.2$\pm$1.5&-33.1&15.5&T\\
 &T3&..&..&22.48$\pm$0.05&702&7.7$\pm$0.6&-40.0&16.1&T\\
A2667&A1&1.034&7&19.67$\pm$0.01,19.33$\pm$0.01,18.26$\pm$0.01&450/606/814&14.6$\pm$2.4&13.4&6.1&T\\
 &A2&..&..&23.07$\pm$0.06,23.59$\pm$0.16,23.50$\pm$0.37&450/606/814&18.1$\pm$8.9&7.2&14.7&T\\
\enddata
\tablenotetext{\dagger}{Photometry and $L/W$ may be affected by poor galaxy subtraction or chip boundary}
\tablenotetext{\ddagger}{Unable to detect feature or get reliable photometry in one of the observed bands}
\tablenotetext{\star}{Redshift is only for feature H1}
\tablenotetext{\ast}{Due to different imaging orientations, this arc only appear in the F702W image}
\tablenotetext{a}{There is a flatfielding problem with this portion of the chip for this image.  While this arc is clearly detected in the F814W band, no attempt was made to correct the flatfielding problem and so no arc magnitude was recoverable.  The $L/W$ for this arc was determined solely from the F555W band.}
\tablerefs{(1) Broadhurst et al. 2000; (2) Smith et al. 2002; (3) Smith et al. 2004; (4) Richard et al. in prep; (5) Sand et al. 2004; (6) Smith et al. 2001; (7) This work; (8) Ellis et al. 1991; (9) Edge et al. 2003; (10) Sand et al. 2002; (11) Mellier et al. 1991; (12) Gladders, Yee, \& Ellingson 2002; (13) Borys et al. 2004; (14) Allen et al. 1996; (15) Holden et al. 2001; (16) PID 6778 abstract; (17) Franx et al. 1997; (18) Yee et al. 1996; (19) Campusano et al. 2001; (20) Smail et al. 1995; (21) Lemoine-Busserolle et al. 2003; (22) Frye \& Broadhurst 1998; (23) Pello et al. 1992; (24) Ebbels et al. 1998; (25) Soucail et al. 1988; (26) Bezecourt et al. 1999; (27) Pello et al. 1999; (29) Tran et al. 2004}
\end{deluxetable}

\end{document}